\documentclass[
aip,
amsmath,
amssymb, 
preprint]{revtex4-1}
\usepackage[english]{babel}
\usepackage{graphicx}% Include figure files
\usepackage{dcolumn}% Align table columns on decimal point
\usepackage{bm}% bold math
\usepackage[utf8]{inputenc}
\usepackage[T1]{fontenc}
\usepackage{mathptmx} 
\usepackage{bbold}
\usepackage{dsfont}
\usepackage{cases}
\usepackage{xcolor}
\usepackage{ulem}
\usepackage{graphicx}
\usepackage{dcolumn}% Align table columns on decimal point
\usepackage{bm}% 
\usepackage{numprint} 
\usepackage[version=3]{mhchem} % Formula subscripts using \ce{}
\usepackage{multirow}
\usepackage{bm}% bold math
\usepackage{booktabs}
\usepackage[round-mode=places,round-precision=2]{siunitx}
\usepackage{blkarray}
\usepackage{color}
\usepackage{morefloats}
\usepackage{enumitem}
\usepackage{blindtext} % to generate dummy text
\usepackage{ltxtable}
\usepackage{setspace}

\usepackage{float}

\usepackage{xr}
\usepackage{soul}
\preprint{MCD}

\begin{document}
\title{Magnetic circular dichroism within the Algebraic Diagrammatic Construction scheme of the polarisation propagator up to third order}
\author{Daniil A. Fedotov}
\affiliation{DTU Chemistry, Technical University of Denmark, Kemitorvet, Building 207, DK-2800 Kongens Lyngby, Denmark}
\author{Mikael Scott}
\affiliation{University of Heidelberg Im Neuenheimer Feld 205A 69120, Heidelberg, Germany}
\author{Maximilian Scheurer}
\affiliation{University of Heidelberg Im Neuenheimer Feld 205A 69120, Heidelberg, Germany}
\author{Dirk R. Rehn}
\affiliation{University of Heidelberg Im Neuenheimer Feld 205A 69120, Heidelberg, Germany}
\author{Andreas Dreuw}
\email{dreuw@uni-heidelberg.de}
\affiliation{University of Heidelberg Im Neuenheimer Feld 205A 69120, Heidelberg, Germany}
\author{Sonia Coriani}
\email{soco@kemi.dtu.dk}
\affiliation{DTU Chemistry, Technical University of Denmark, Kemitorvet, Building 207, DK-2800 Kongens Lyngby, Denmark}

\begin{abstract}
We present an implementation of the $\mathcal{B}$ term of Magnetic Circular Dichroism within the Algebraic Diagrammatic Construction (ADC) scheme of the polarization propagator and its Intermediate State Representation.
As illustrative results, the MCD %B term 
spectra of the ADC variants ADC(2), ADC(2)-x and ADC(3) of the molecular systems uracil, 2-thiouracil, 4-thiouracil, purine, hypoxanthine 
1,4-naphthoquinone, 9,10-anthraquinone and 1-naphthylamine , are computed and compared with results obtained using the Coupled-Cluster Singles and Approximate Doubles (CC2) method, {literature TD-DFT results} as well as with available experimental data. 
\end{abstract}


\maketitle

\section{Introduction}
\label{Intro}
% \revD{
%designation
Magnetic Optical Activity (MOA) collectively indicates the nonlinear optical effects that arise when electromagnetic radiation impinges matter in presence of a magnetic field longitudinal to the direction of propagation of light.~\cite{BuckinghamMCD,Barron:2004,CPP:MCHD} 
Among a number of MOA effects,  Magnetic Optical Rotation (MOR) -- also known as Faraday rotation -- and Magnetic Circular Dichroism (MCD) are the prominent ones.~\cite{Serber.PhysRev.41.489,BuckinghamMCD,Barron:2004} MOR originates from the difference in the refractive indices observed when irradiating with either right or left circularly polarized light, whereas MCD comes from the corresponding difference in absorption coefficients.~\cite{Serber.PhysRev.41.489,BuckinghamMCD,Stephens:MCD:1970,Stephens:MCD:1974,MCD.ch4,Barron:2004} 
Both phenomena have found many applications in physics, chemistry and spectroscopy.
%IDENTIFY BETTER REFERENCES
%~\cite{Cukras2016,MCD:theory:Bio,MCD:organicmol,MCD:organicmolec}
% B
%
% Absorptive component of MOA - Magnetic Circular Dichroism (MCD) causes from difference in absorption of left and right circularly polarised light. MCD phenomena have found many applications in physics, chemistry and spectroscopy.~\cite{Barron:2004,MCD_Mason_book,Cukras2016,MCD:theory:Bio,MCD:organicmol,MCD:organicmolec} 
In particular, MCD is at the foundation of the spectroscopy bearing the same name, which is a
popular technique, 
complementary to conventional one-photon absorption, to study the geometric, electronic, and magnetic properties of chemical systems,
in gas phase, in solution, or as isotropic solids.~\cite{MCD_book_Mason} 
Since the MCD spectral features are signed
and depend on the magnetic moments of the molecular electronic states, as well as on the direction of the field, 
MCD yields additional information and can reveal hidden electronic transitions. 
MCD spectra can be equally recorded for achiral and chiral samples.

% Electronic Structure (ES) investigation of organic molecules and compounds because
% magnetically perturbed ES exhibits transitions which were not allowed in One-Photon Absorption (OPA) or Electronic Circular Dichroism (ECD) spectra.\cite{MCD:theory:Bio,ur_exp_vapor,OPA+MCD_exp}. 

% The spectral signatures of a MCD spectrum were already in the sixties, \citeauthor{BuckinghamMCD}\cite{BuckinghamMCD}
% presented a detailed theoretical treatment of the dispersion of the Faraday effect through absorption bands,
% presenting Sum-Over-State (SOS) expressions for MCD was offered by~\citeauthor{BuckinghamMCD} in 1966 and can be divided on three terms: A, B and C. A and C terms are not equal zero only in the presence of degenerate excited states and for open-shell systems, respectively. B term has dominant particular importance for, so in this study, we are focused on the B term of MCD.\cite{Coriani:MCD:2008}
The MCD spectral bands are traditionally characterized in terms of three Faraday terms,~\cite{Serber.PhysRev.41.489,MCD:theory:Bio,MCD_book_Mason} called $\mathcal{A}$, $\mathcal{B}$ and $\mathcal{C}$. For organic molecules, the $\mathcal{B}$ term is of special interest, since the 
$\mathcal{A}$ term only contributes if either the ground state or the final excited state are degenerate, while the $\mathcal{C}$ term only contributes if the ground state possesses a permanent magnetic moment (degenerate ground state). 
According to perturbation theory,~\cite{BuckinghamMCD,Stephens:MCD:1970,Stephens:MCD:1974,MCD.ch4,MCD_2012} the $\mathcal{B}$ term describes the electric dipole transition from the initial 
to the  final state 
through 
%
%\revD{is determined by transition electric dipole moment between the two states (excited or ground states) and transition magnetic dipole moment between ground and excited states
the coupling by the magnetic dipole moment
with a set of intermediate excited states
(see Eq.~\eqref{SOS_MCD_STRESP}).

Alike other spectroscopies, the detailed interpretation and assignment of the MCD spectral features relies on theoretical simulations. The first theoretical approaches to compute MCD spectra were fundamentally based on the sum-over-state (SOS) expressions~\cite{SOS:MCD:1972,SOS:MCD:1973,SOS:MCD:1980,SOS:MCD:1974} of the
three MCD terms,
and date back to the seventies and the beginning of the eighties. SOS-based methods require the explicit calculation of a large number of excited states as well as electronic and magnetic dipole transition moments between them. 
One-by-one consideration of all intermediate states clearly limits the applicability of SOS methods, 
%to relatively small systems, 
especially when the $\mathcal{B}$ term is of interest.
{A more efficient strategy is offered by response theory}, where the summation over explicitly determined intermediate states is replaced
by the solution of 
linear response equations.~\cite{Response1,MCD_Coriani_1999}
This is the approach followed in many implementations of the calculation of the MCD $\mathcal{A}$ and $\mathcal{B}$ terms presented the last two decades at different levels of theory, including
Self-Consistent Field (SCF) and Multi-Configurational Self-Consistent Field,~\cite{MCD_Coriani_1999}
Time-Dependent Density Functional Theory (TD-DFT),~\cite{Seth2004,Seth2008:b,Seth2008:a,Solheim2008:b,Seth2010,MCD:DFT:Coriani}
and Coupled Cluster (CC) theory.~\cite{MCD_Coriani_2000,Sarah:MCD,MCD:CC:2020} 
{Alternative to the explicit determination of the $\mathcal{A}$, $\mathcal{B}$ and $\mathcal{C}$ terms, in damped response theory~\cite{Norman:CPP:2001,Norman:PCCP:2011} one computes directly the
MCD cross section on a grid of frequencies, which may prove particularly convenient for systems with a high density of states.~\cite{Krykunov2007,CPP:MCD,MCD:CC:2020}}
A special mention also goes to the approach  by~\citeauthor{Neese_MCD},~\cite{Neese_MCD} where the MCD spectra are obtained directly from the difference between
left and right circularly polarized transition
probabilities in the presence of an external magnetic
field. This involves the calculation of transition energies and intensities using ground- and excited-state wave functions perturbed by the spin–orbit coupling and Zeeman interactions.
% where MCD transitions are computed by the explicit treatment of spin-orbit coupled and spin-spin coupled states of N-electron system what makes possible to calculate simultaneously all MCD terms.}
Real-time approaches are also available to simulate MCD spectra, in particular at TD-DFT level.~\cite{RT-MCD,MCD:Xiaosong:2019}

% \revS{remember to add a sentence on Neese's approach }
%I am not sure what you meant with the sentence below
%This sentence is broken. The sentence was about possibility to apply successfully SOS methods for MCD calculation, if one knows states with highest contribution.    
% organic systems without comprehension information of the nature of excited states~\cite{MCD_2012}. 

% In the last few decades MCD spectroscopy has been under strong interest from both methodological and applicative sides what initiated an active development of new theoretical approaches based on Response Theory~\cite{CPP:MCD:PE,DampedResponse} (RT):
% Self-Consistent Field (SCF) and Multi-Configurational Self-Consistent Field~\cite{MCD_Coriani_1999} (MCSCF), Coupled Cluster~\cite{MCD_Coriani_2000,Sarah:MCD} (CC), Density Functional Theory~\cite{Seth2004,Seth2008:b,Seth2008:a,Solheim2008:b,Seth2010,MCD:DFT:Coriani} approaches. 

The purpose of this study is to extend the portfolio of spectroscopies that can be simulated using the Algebraic Diagrammatic Construction (ADC) scheme of the polarization propagator~\cite{ADCden}  to include MCD.
First introduced, in its second-order variant ADC(2), by Schirmer in 1984,~\cite{ADC:orig} the ADC approach has the following two decades~\cite{ADC3:1999,ADC3compareFCICC} found its primary applications to the investigation of photo-ionization processes~\cite{ADC_NonDyson,HANDKE1995109} and to X-ray spectroscopy.~\cite{ADC:Barth1985,ADC:Plekan2008,ADC:Trofimov2000,ADC:CVS} Following the introduction of the intermediate state representation,~\cite{ISR_PhysRevA.43.4647,SchirmTrofADC}
%in 2004,~\cite{ISR_PhysRevA.43.4647,SchirmTrofADC} 
which has significantly simplified the derivation of computational expressions for other types of molecular properties and spectroscopy,
the ADC family of methods has been steadily
gaining popularity within the computational chemistry community. A fundamental contribution in this regard has been its implementation within one of the leading
quantum chemistry software platforms.~\cite{ADC:Xray,ADC3:2014,Qchem_MP_paper,Qchem541}
To date, 
beside the already mentioned ionization processes, X-ray absorption and photoemission,
it can be used to compute, e.g. 
two-photon absorption,~\cite{ADC_TPA}
resonant inelastic X-ray scattering,~\cite{CPP:RIXS:ADC}
electric dipole polarizabilities of ground~\cite{CPP:ADC}
and excited states,~\cite{scheurer2020complex}
hyperpolarizabilities, and most recently electronic circular dichroism.~\cite{ADC:ECD:Scott2021}
%
% For this aim, we employ the RT approach for derivation implementation of a programmable expression for MCD $\mathcal{B}$ term for ADC wave function. 
% Another \textit{ab initio} approach which demonstrates high potential for simulation of MCD spectroscopy is Algebraic Diagrammatic Construction method (ADC).~\cite{ADCden} 
%\revD{
% The ADC family of methods
% has been gaining popularity over the last two decades in many areas of spectroscopy and has shown great success in investigating photo-ionization processes~\cite{!ref!}, linear and non-linear effects in X-ray region.~\cite{CPP:RIXS:ADC,ADC:CVS,ADC:Xray} 
The ADC methods are often compared to~\cite{ADCandCC:Haettig,HATTIG200537,RICC2TMES,ADCfamily:Dreuw} coupled cluster response methods~\cite{CCRF,Christiansen:1998} and equation of motion coupled cluster.~\cite{Stanton:93:EOMCC,coriani-cc-ci}
More recently, they have been
put in close relation with
the unitary coupled cluster ansatz.~\cite{Liu:UCC3:ADC3,Hodeker:2020:UCC,Hodecker:ADCvsCC}
%one but has some important advantages. 
Different from ``traditional'' (projection-based) CC, however, 
the ADC secular matrix (the analogue of the coupled cluster Jacobian)
is Hermitian,
which simplifies the
calculation of transition properties, since one does not need to compute both right and left excitation vectors and transition moments.
Compared to CC response theory,
the ADC response functions do not contain additional non-linear terms. 
Moreover, the ground state of ADC($n$) is 
%typically 
%identified with second-order 
given by the wave function and energy of M{\o}ller-Plesset perturbation theory (MP$n$),
which is non-iteratively determined 
(in contrast to the iterative CC wave functions).
As result, the ADC($n$) methods are 
computationally cheaper than the corresponding 
%(as far accuracy is concerned) 
CC approximations of same formal scaling with the system size. 
The paper is organized as follows. 
The reformulation of the MCD $\mathcal{B}$ term expression in the ADC/ISR formalism, together with details on implementation are presented is Section~\ref{theory}. The computational details of the numerical studies are described in Section~\ref{CompDet}. In Section~\ref{Results} we present and discussion our results for a number of 
test systems. Some final remarks are given in Section~\ref{Conclusions}.

\section{Theory and Methodology}
\label{theory}
\subsection{The MCD $\mathcal{B}$ term in the ADC/ISR formalism}
A spectral (i.e., SOS) representation of the MCD $\mathcal{B}$ term for a transition $0\to j$ was derived using perturbation theory by Buckingham and Stephens~\cite{BuckinghamMCD}
\begin{align}
\label{SOS_MCD_STRESP}
{\mathcal{B}} (0 \rightarrow{j})= \Im \Bigg\{ \sum_{k\neq 0}    & \frac{\langle  k|{\hat{\boldsymbol{m}}}| 0 \rangle}{\omega_{k}} \cdot { \langle 0|{\hat{\boldsymbol{\mu}}}| j \rangle}  \times {\langle j|{\hat{\boldsymbol{\mu}}}|k \rangle} 
\\ \nonumber &
+ \sum_{k\neq j} \frac{ \langle j|{\hat{\boldsymbol{m}}}| k \rangle}{\omega_{k} - \omega_{j}} \cdot { \langle 0|{\hat{\boldsymbol{\mu}}}|j \rangle}  \times {\langle k | {\hat{\boldsymbol{\mu}}} | 0\rangle}\Bigg\},
\end{align} 
where, in atomic units, $\hat{\boldsymbol{\mu}} = -\sum_{i} \mathbf{r}_{i}-\sum_{I}Z_{I}\mathbf{R}_{I}$ is the electric dipole moment,
$\hat{\boldsymbol{m}}= -\frac{1}{2}\sum_{i}$ $\mathbf{{r}}_{i} \times \mathbf{{p}}_{i} + \sum_{I}  (Z_{I}/ 2M_{I}) \mathbf{R}_{I} \times \mathbf{P}_{I}$ is the magnetic dipole moment; $\mathbf{r}_{i}$ and $\mathbf{p}_{i}$ are the position and linear momentum of the $i^{th}$ electron; 
$Z_{I}$, $M_{I}$, $\textbf{R}_{I}$ , and $\textbf{P}_{I}$ are the charge, mass, position, and linear momentum of the $I^{th}$ nucleus. %$\operatorname{Im}$
The symbol $\Im$ denotes that the imaginary part of the quantity within curly brackets is to be used. The energies and wave functions are those of the unperturbed system.
{As we will consider only electronic excited states in the following, the nuclear components of the electric and magnetic dipole operators do not contribute to the $\mathcal{B}$ term, so they can be disregarded.}
It is convenient to rewrite
Eq.~\eqref{SOS_MCD_STRESP}
in compact tensor notation~\cite{MCD_Coriani_1999}
\begin{align}
\label{tensor_notation}
{\mathcal{B}} (0 \rightarrow{j})= - \epsilon_{\alpha\beta\gamma}
%\operatorname{Im} 
\Im
\Bigg[ \sum_{k\neq 0}   \frac{\langle 0|{\hat{m}}_{\beta}| k \rangle { \langle k | {{\hat{\mu}}}_{\gamma} | j \rangle }}{\omega_{k}}   
+  \sum_{k\neq j} \frac{ \langle k|{{{\hat{m}}}}_{\beta}| j \rangle {\langle 0 | {{\hat{\mu}}_{\gamma}} | k\rangle}}{\omega_{k} - \omega_{j}}  \Bigg]\langle j|{\hat{\mu}}_{\alpha}| 0 \rangle
\end{align} 
where Greek indices correspond to specific Cartesian components, and $\epsilon_{\alpha\beta\gamma}$  is the Levi-Civita tensor. Implicit summation over repeated indices is assumed.
% \citeauthor{MCD_Coriani_1999}\cite{MCD_Coriani_1999}
% showed that it can be connected to the single residue of the dipole-dipole-magnetic dipole quadratic response function 
% %
% \begin{equation}
% \label{QR_STRSP_1}
% % B_{i,j} = 
% % \operatorname{Im} \Bigg[ \epsilon_{\alpha\beta\gamma} \lim_{\omega\to\omega_{j}}(\omega-\omega_{j}) \langle\langle {\mu_{\gamma};m_{\beta},\mu_{\alpha} \rangle\rangle}^{\gamma}_{\omega,0} \Bigg]
% B (0 \rightarrow{j})= \epsilon_{\alpha\beta\gamma}
% \operatorname{Im} \Bigg[  \lim_{\omega\to\omega_{j}}(\omega-\omega_{j}) \langle\langle {\mu_{\gamma};m_{\beta},\mu_{\alpha} \rangle\rangle}_{0,\omega} \Bigg]
% \end{equation} 
% %\begin{equation}
% %\label{QR_STRSP_1}
% %\lim_{\omega\to\omega_{j}}(\omega-\omega_{j}) \langle\langle {\mu_{\gamma};m_{\beta},\mu_{\alpha} \rangle\rangle}_{\omega,0} 
% %\end{equation} 
Eq.~\eqref{tensor_notation}
has earlier been shown to be connected to the single residue of the dipole-dipole-magnetic dipole quadratic response function.~\cite{Response1,MCD_Coriani_1999} 
For our subsequent formulation within ADC/ISR,
it is advantageous to extract the $k=0$ contribution from the second summation and add it to the first one, which yields
\begin{align}
\label{prova_C}
{\mathcal{B}} (0 \rightarrow{j}) 
= - \epsilon_{\alpha\beta\gamma}
\Im & \Bigg\{  \Bigg[
\sum_{k\ne 0}
\frac{\langle 0|{\hat{m}}_{\beta}| k \rangle 
{ \langle k | \hat{\bar{\mu}}_\gamma | 
%0\rangle)
 j \rangle }}{\omega_{k}}
% { \langle k | {({\hat{{\mu}}}}_{\gamma} - \langle 0 |\hat{\mu}_\gamma | 0\rangle)| j \rangle }}{\omega_{k}}
% + 
% \frac{\langle 0|{\hat{m}}_{\beta}| j \rangle { \langle j | {{\hat{{\mu}}}}_{\gamma} | j \rangle }}{\omega_{j}}
\\ \nonumber
& + \sum_{k\neq \{0,j\}} \frac{ \langle k|{{\hat{{m}}}}_{\beta}| j \rangle {\langle 0 | {{\hat{\mu}}_{\gamma}} | k\rangle}}{\omega_{k} - \omega_{j}} 
\Bigg]
\langle j|{\hat{\mu}}_{\alpha}| 0 \rangle
\Bigg\}
\end{align} 
In the first summation, the dipole transition
moments between states $k$ and $j$ are computed using the fluctuation dipole operator $\hat{\bar{\mu}}_{\gamma}$ = $\hat{\mu}_{\gamma}$ -- $\langle0|\hat{\mu}_{\gamma}|0\rangle$. In the second summation, 
%using 
the ground state configuration of the fluctuation magnetic dipole operator is omitted, since $\langle 0| m| 0\rangle=0$ for non-degenerate ground states (as considered in this study).
To obtain the ISR form of the
$\mathcal{B}$ term
we translate the transition moments and energy denominators in Eq.~\eqref{prova_C} in terms of ADC/ISR fundamental building blocks.~\cite{ADCfamily:Dreuw,scheurer2020complex}
Thus, the transition moment between the ground state $\Psi_0$ and an excited state $\Psi_n$ for a generic operator $\hat{O}$
is written as
\begin{equation}
\label{Tn}
{T}_{0n}({\hat{O}}) \equiv \langle {\Psi}_{0}|{\hat{O}}| \Psi_{n}\rangle = \mathbf{F}^{\dagger}({\hat{O}}) Y_{n}~. 
\end{equation}
The transition moment between two excited states for the (fluctuation) operator $\hat{\bar{O}}$
is obtained as
\begin{equation}
\label{Tnm}
{T}_{nm}({\hat{\bar{O}}}) \equiv \langle {\Psi}_{n}|\hat{\bar{O}}| \Psi_{m}\rangle = Y^{\dagger}_{n} \mathbf{B}({\hat{\bar{O}}}) Y_{m} 
\end{equation} 
In the expressions above, 
$\mathbf{F}$ and $\mathbf{B}$ are, respectively, a vector and a matrix, whose elements are 
defined in the ISR basis~$\{\tilde{\Psi}_I\}$
\begin{equation}
\label{F}
F_{I}({\hat{O}}) 
= \langle \widetilde{\Psi}_{I}|{\hat{O}}|\Psi_{0} \rangle 
= \sum_{pq} O_{pq} f_{I,pq}
\end{equation}
and 
\begin{equation}
\label{B}
% B_{IJ}({\hat{\bar{O}}})  = \langle \widetilde{\Psi}_{I}|\hat{O}-{\hat{O}_{0}} |\widetilde{\Psi}_{J} \rangle.
B_{IJ}({\hat{\bar{O}}})  = \langle \widetilde{\Psi}_{I}|\hat{O} |\widetilde{\Psi}_{J}\rangle
-\delta_{IJ} \langle {\Psi}_{0}|\hat{O}|{\Psi}_{0}
\rangle.
\end{equation}
The intermediate state basis $\{\tilde{\Psi}_I\}$
is generated by 
Gram-Schmidt orthogonalization of a set 
of precursor states
%excited state Slater determinants  
$\{{\Psi^{I}_{0}}^{\#}\}$,
obtained by acting on the 
formally exact
ground state $|{\Psi}_0\rangle$
with the excitation operators $\{\hat{C_I}\}$,
% \begin{equation}
% \label{C}
% $\{ \hat{C}_I\} \equiv \{ \hat{c}^{\dagger}_{a}\hat{c}_{i},\hat{c}^{\dagger}_{a}\hat{c}^{\dagger}_{b}\hat{c}_{i}\hat{c}_{j},\cdots \}$,  ($i<j$, $a<b$, $\cdots$),
%\end{equation}
\begin{equation}
\label{IS}
\hat{C}_{I} | \Psi_{0}\rangle = |{\Psi^{I}_{0}}^{\#}\rangle \xrightarrow{\text{Gram-Schmidt}} | \widetilde{\Psi}_{I}  \rangle. 
\end{equation}
% by the action of some excitation operators $\{ \hat{C_I}\}$
% \begin{equation}
% \label{C}
% \{ \hat{C}_I\} \equiv \{ \hat{c}^{\dagger}_{a}\hat{c}_{i},\hat{c}^{\dagger}_{a}\hat{c}^{\dagger}_{b}\hat{c}_{i}\hat{c}_{j},\cdots \},  (i<j, a<b, \dots)
% \end{equation}
% Following standard notation, indices $a, b, \cdots$ refer to virtual orbitals and $i, j, \cdots$ to occupied ones.

In Eqs~\eqref{Tn} and \eqref{Tnm}, $Y_n$ and $Y_m$ are the $n^{th}$
and $m^{th}$ eigenvectors obtained diagonalizing the Hermitian ADC matrix $\mathbf{M}$,
\begin{equation}
\mathbf{M} \mathbf{Y}=  \mathbf{Y} \boldsymbol{\Omega}, \quad \mathbf{Y}^\dagger \mathbf{Y} = \mathbf{1}.
\label{myeqomy}
\end{equation} 
$\mathbf{M}$ is also defined in the ISR basis as
\begin{align}
\label{MandFinISR}
{M}_{{I} {J}} &=  \langle \widetilde{\Psi}_{I} |\hat{H} - {E_{0}}| \widetilde{\Psi}_{J}  \rangle, 
\end{align}
and
\begin{align}
\label{omega}
 {\Omega}_{{n} {m}}&=  \langle {\Psi}_{n} |\hat{H} - {E_{0}}| {\Psi}_{m}  \rangle,
\end{align}
is the diagonal matrix of the eigenvalues.
In other words, $\mathbf{Y}=\{Y_n\}$ is the unitary matrix that connects the 
(diagonal) basis of {exact} excited states $\{ \Psi_n\}$ with the 
%(primitive)
ISR basis, according to~\cite{ADC:orig}
%standard linear algebra~\cite{helgaker:pinkbook}
\begin{align}
\label{inter}
|{\Psi}_{n} \rangle = \sum_{J} Y_{Jn} |\widetilde{\Psi}_{J} \rangle, \quad Y_{Jn} =\langle \widetilde{\Psi}_{J}|\Psi_{n}
\rangle.
\end{align} 

With the above building blocks, the ADC/ISR expression of the $\mathcal{B}$ term becomes
\begin{align}
\label{BinADC}
% B (0\rightarrow{j}) = \epsilon_{\alpha\beta\gamma}
% \Bigg[ & F^{\dagger}({\hat{m}_\beta})\mathbf{M}^{-1}B(\hat{\mu}_\gamma){Y}_{j} + \frac{{m}^{0j}_{\beta}{\mu}^{jj}_{\gamma}}{\omega_{j}} +\\ & F^{\dagger}({\hat{\mu}_\gamma})(\mathbf{M}-\omega_j \mathbf{1})^{-1}B(\hat{m}_{\beta}) {Y}_{j} +  \frac{{m}^{0j}_{\gamma}{\mu}^{00}_{\beta}}{-\omega_{j}} \Bigg] \mathbf{F}({\hat{\mu}_\alpha} ){Y}_{j},
{\mathcal{B}} (0\rightarrow{j}) = \epsilon_{\alpha\beta\gamma}
& \Bigg[\mathbf{F}^{\dagger}({\hat{m}_\beta}) %YY^{\dagger}
\mathbf{M}^{-1} %YY^{\dagger}
{\mathbf{B}}(\hat{{\mu}}_\gamma){Y}_{j} 
\\\nonumber
& + \mathbf{F}^{\dagger}({\hat{\mu}_\gamma})(\mathbf{M}-\omega_j \mathbf{1})^{-1}{\mathbf{B}}(\hat{\bar{m}}_{\beta}) {Y}_{j} 
% \\
% + \frac{{m}^{0j}_{\beta}({\mu}^{jj}_{\gamma}-{\mu}^{00}_{\gamma})}{\omega_{j}} 
\Bigg] {Y}^\dagger_{j}\mathbf{F}({\hat{\mu}_\alpha})~,
%\alert{{Y}^\dagger_{j}\mathbf{F}({\hat{\mu}_\alpha}) ?}
\end{align} 
{where the condition $k\ne j$ in the second SOS contribution (Eq.~\eqref{SOS_MCD_STRESP}) is enforced using a projector that removes $j$ from the response equations (vide infra).} 
% where $m^{0j}_\beta=\langle 0 | \hat{m}_\beta|j\rangle$ and similarly for the other terms. 
% \revD{The last term in square brackets  included in Eq.~\eqref{BinADC}}   
% %$\frac{{m}^{0j}_{\gamma}{\mu}^{00}_{\beta}}{-\omega_{j}}$ and $\frac{{m}^{0j}_{\beta}{\mu}^{jj}_{\gamma}}{\omega_{j}}$ was explicitly 
% \alert{what does this mean:} 
% \revD{because of the ground state shift in the definition of matrix $\mathbf{B}$, Eq.~\eqref{B}}\alert{???}.\\
% % \alert{I AM STILL NOT SURE WHY WE HAVE A JJ TERM AND 
% % WHAT THE SIGNS ARE. IN THE CODE IT LOOKS LIKE 
% % $(\mu^{jj}-\mu^{00})$ (FOR THE SAME DIPOLE COMPONENT), BUT I CANNOT SEE HOW THIS CORRESPONDS TO "ADDING BACK" THE 00 CONTRIBUTION OF THE FLUCTUATION (SHOULD'T IT BE A +?)
% % NOR WHERE THE JJ COMES FROM...}

\subsection{Implementation}
To compute the $\mathcal{B}$ term according to Eq.~\eqref{BinADC}, the response vectors $X^{\hat{m}}$ and $X^{\hat{\mu}}$ are evaluated solving the following response equations: 
\begin{equation}
\label{X_m}
\mathbf{M} X^{\hat{m}_\beta} = \mathbf{F}(\hat{m}_{\beta}) 
\end{equation} 
and 
\begin{equation}
\label{X_mu}
(\mathbf{M}-\omega_j \mathbf{1}) X^{\hat{\mu}_\beta} = \mathbf{F}(\hat{\mu}_{\beta}).  
\end{equation}
using a Jacobi algorithm with Direct Inversion of the Iterative Subspace (DIIS)~\cite{DIIS} acceleration. The equations are solved for all three Cartesian components of the electric dipole and magnetic dipole operators. 
As seen, Eq.~\eqref{X_mu} also contains an explicit dependence on the excitation energy $\omega_j$ of the final excited state $j$. 
Therefore it is potentially divergent if the solution vector $X^{\hat{\mu}}$ contains a component along the direction of $Y_j$.
However, the restriction in the summation of the second term of Eq.~\eqref{BinADC}, ($k \ne j$), imposes that the solution of Eq.~\eqref{X_mu} is found in the space orthogonal to the eigenvector $Y_{j}$. This is done by applying a Gram-Schmidt orthogonalization procedure during the solution of the response equation.
% \alert{Was the projection applied also to Eq.~\eqref{X_m} for the magnetic moment? That would explain the addition of the last term in the $\mathcal{B}$ formula}
%
The number of response equations 
%to be solved 
is equal to $3 \times (N_j + 1)$, where $N_j$ is the number of final excited states.

The final ADC computational expression thus reads
% (\revS{maybe also translate the $m^{0j}(\mu^{jj}-\mu^{00})$ terms in building blocks?})
\begin{equation}
\label{BinADC_impl}
% B (0\rightarrow{j}) = \epsilon_{\alpha\beta\gamma}
% \Bigg[ & F^{\dagger}({\hat{m}_\beta})\mathbf{M}^{-1}B(\hat{\mu}_\gamma){Y}_{j} + \frac{{m}^{0j}_{\beta}{\mu}^{jj}_{\gamma}}{\omega_{j}} +\\ & F^{\dagger}({\hat{\mu}_\gamma})(\mathbf{M}-\omega_j \mathbf{1})^{-1}B(\hat{m}_{\beta}) {Y}_{j} +  \frac{{m}^{0j}_{\gamma}{\mu}^{00}_{\beta}}{\omega_{j}} \Bigg] \mathbf{F}({\hat{\mu}_\alpha} ){Y}_{j},
{\mathcal{B}} (0\rightarrow{j}) = - \epsilon_{\alpha\beta\gamma}
\Bigg[X^{m_\beta \dagger}
{\mathbf{B}}(\hat{{\mu}}_\gamma){Y}_{j} +   X^{\mu_\gamma\dagger}{\mathbf{B}}(\hat{m}_{\beta}) {Y}_{j} 
% + \frac{{m}^{0j}_{\beta}({\mu}^{jj}_{\gamma}-{\mu}^{00}_{\gamma})}{\omega_{j}} 
\Bigg] {Y}^\dagger_{j} \mathbf{F}({\hat{\mu}_\alpha})~.
\end{equation} 
\section{Computational Details}
\label{CompDet}
\begin{figure}[hbt!]
\centering
\includegraphics[width=0.7\linewidth]{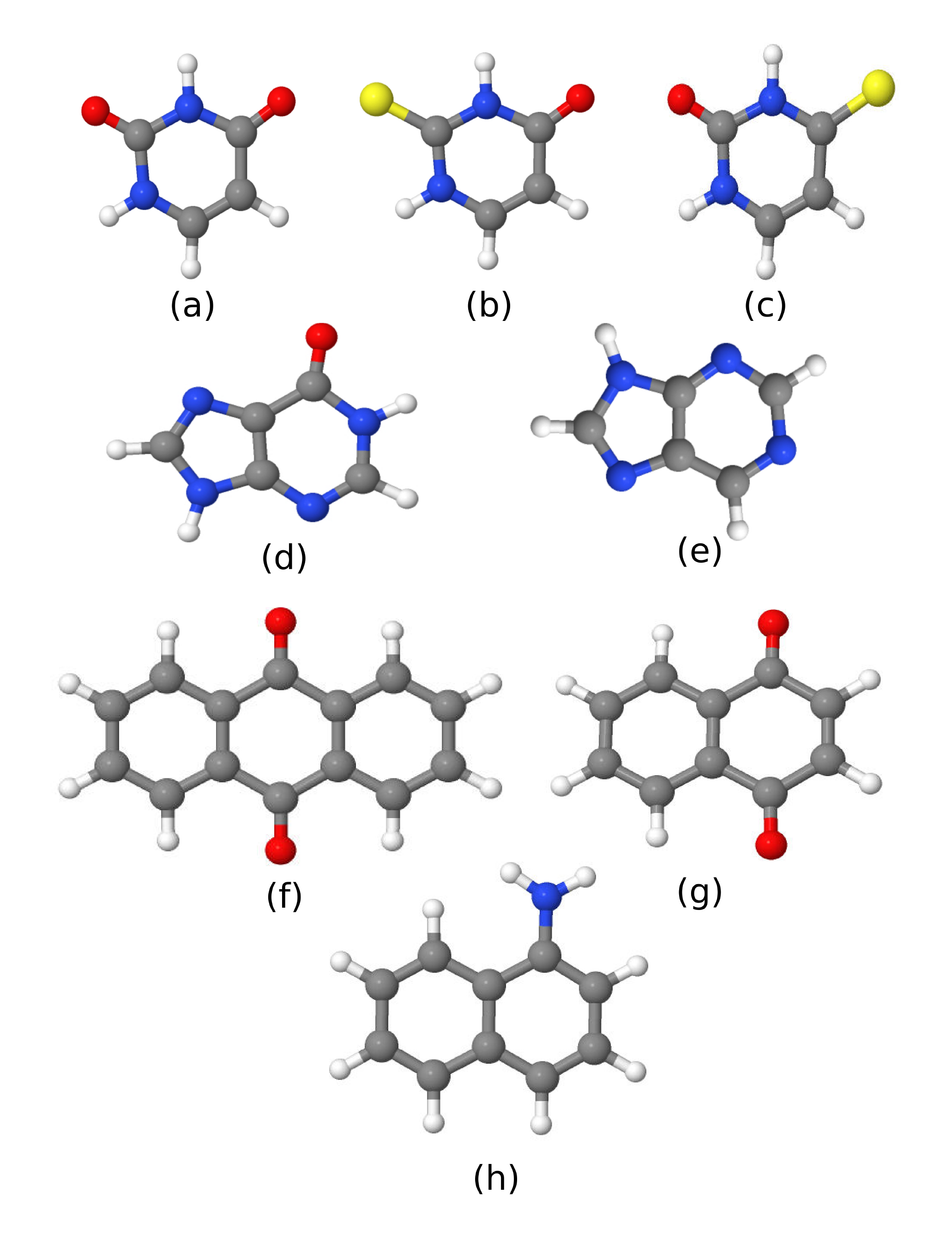}
\caption{The molecules considered in this study: (a)~uracil, 
(b)~2-thiouracil, 
(c)~4-thiouracil, 
(d)~hypoxanthine, 
(e)~purine, 
(f)~9,10-anthraquinone, 
(g)~1,4-naphthaquinone,
(h)~1-naphthylamine.}
\label{allmols}
\end{figure}

\begin{figure}[hbt!]
\centering
\includegraphics[width=0.9\linewidth]{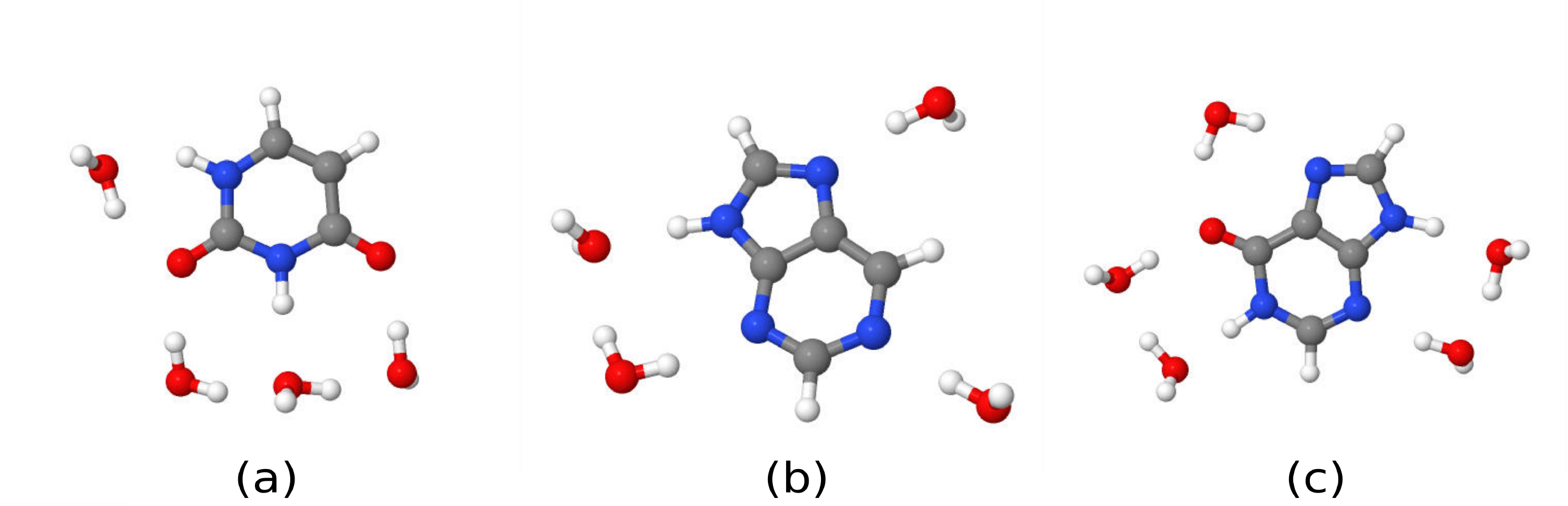}
\caption{The clusters with water considered in this study: 
(a)~uracil + 4$\cdot$H$_2$O; (b)~purine + 4$\cdot$H$_2$O, (c)~hypoxanthine + 5$\cdot$H$_2$O. \label{allmols+w}}
\end{figure}

The MCD 
%calculation of the 
$\mathcal{B}$ term for the ADC(2), ADC(2)-x, and ADC({3}) methods was implemented 
using the ISR at second order
in a development version of Q-Chem.~\cite{Qchem_MP_paper,Qchem541} 
{The methodology developed was tested on different molecular systems: uracil, 2- and 4-thiouracil, purine, hypoxanthine, 1,4-naphthoquinone, 9,10-anthraquinone and 1-naphthylamine.}
{We will focus our discussion primarily on the ADC(2) and ADC(3) results, while the ADC(2)-x data is given in the SI for completeness.}
{The molecular structures (see Fig.~\ref{allmols}) of uracil, purine, hypoxanthine were optimised at the MP2/cc-pVTZ level, whereas those of the thiouracils and  of 9,10-anthraquinone  at the B3LYP/cc-pVTZ level of theory. All optimizations were carried out using Q-Chem.} To estimate the solvent effects, we also considered clusters of purine, uracil and hypoxanthine with explicit water 
molecules, whose structures were taken from a previous computational study.~\cite{CPP:MCD:Nucleic}
They are shown in Figure~\ref{allmols+w}.
% \revS{Which are MP2 and which B3LYP? Did you optimize all of them, or are some of them taken from the literature?}
% \revD{Stuructures were obtained: 
% uracil(the one we used in uracil paper),
% 2-thiouracil (from Shota),
% purine and hypoxanthine (obtained by me)
% 4-thiouracil (from Sonia), 
% 1-Naphthylamine (from Simone),
% 1,4-naphtaquinone (from Xiaosong, the one used in the paper (Sun, et al. 2019)
% 9,10-anthraquinone (obtained by me)
% }
% \revD{The structures of 2-thiouracil and 1-Naphthylamine were optimised at RI-MP2/cc-pVTZ and RICC2/aug-cc-pVDZ levels of theory using Turbomole.~\cite{Turbomole2020} The structure for 1,4-naphtaquinone was from \citenumber{MCD:Xiaosong:2019}}

The correlation-consistent aug-cc-pVDZ basis set~\cite{Dunning_basis}  was used in the  calculations of the $\mathcal{B}$ term. 
The RICC2 calculations were carried out with Turbomole.~\cite{Turbomole2020}
{In all calculations the gauge origin was placed in center of the nuclear charges.}
The convergence threshold for response equations was set equal to 10$^{-4}$ 
%and 10$^{-5}$ 
in all calculations. 
{For ease of comparison with the experimental bands, we report in the tables the negative of the $\mathcal{B}$ term---according to the standard convention, a positive MCD band corresponds to a negative $\mathcal{B}$ term.~\cite{MCD_book_Mason}}
 %The OPA and B term MCD spectra are presented in the decadic molar extinction coefficients $\epsilon$, and the extinction coefficient anisotropy, $\Delta\epsilon$, respectively:
The OPA spectra are reported as decadic molar extinction coefficients $\epsilon (\omega)$ (in standard units M$^{-1}$cm$^{-1}$):~\cite{rizzo2012}
\begin{equation}
\epsilon(\omega) = 703.301 \times \omega \times \sum_j g(\omega,\omega_j) \frac{ 3 f_j}{2\omega_j} 
\end{equation}
where $\omega_j$ (in a.u.) and $f_j$ are the excitation energy and oscillator strength  of electronic transition $j$, respectively, $\omega$ (in a.u.) is the frequency of the incoming radiation, and $g(\omega,\omega_j)$ is a broadening function.

The MCD spectra are reported as extinction coefficient anisotropy, $\Delta\epsilon$ (in units M$^{-1}$cm$^{-1}$T$^{-1}$), which are connected to the 
$\mathcal{B}$ terms by~\cite{rizzo2012} 
\begin{equation}
\Delta\epsilon = - \frac{8\pi^2N_A\omega {B}_\textrm{ext}}{3\times1000\times \ln(10)(4\pi\epsilon_{0}\hbar c_0)} \sum_j g(\omega, \omega_j) {\mathcal{B}}(0\rightarrow{j}),
\end{equation}
where ${B}_\textrm{ext}$ is the strength of external magnetic field, $\epsilon_0$ is the vacuum permittivity, and $N_A$ is Avogadro's number.
In both expressions, we use a Lorentzian broadening function 
\begin{equation}
\label{broadening}
g(\omega,\omega_j) = \frac{1}{\pi}\frac{\gamma}{ (\omega-\omega_j)^2 + \gamma^2 }~,
\end{equation}
with a half width at half maximum (HWHM) value of $\gamma~=~0.0045563$  Hartree (1000 cm$^{-1}$), if not stated otherwise.

% Experimental spectra for 1,4-naphthoquinone and 9,10-anthraquinone, were re-digitized from Ref.~\citenum{exp:Anthra:Naphta}, and scaled by a factor 10$^{4}$, %\revS{I STILL DO NOT UNDERSTAND THE NEXT SENTENCE!}
% as we believe that the experimental data was normalized to 1 Gauss. This assumption is supported by the experimental results reported by~\citeauthor{910Anthra_exp2}~\cite{910Anthra_exp2} for 9,10-anthraquinone. 
%
% SONIA I am taking all this out.
%
% \revD{Experimental MCD spectra for 1,4-naphthoquinone and 9,10-anthraquinone were re-digitized from Ref.~\citenum{exp:Anthra:Naphta}. 
% They were there given as [$\theta$]$_{\text{M}}$.
% In order to convert to $\Delta\epsilon$ in standard units M$^{-1}$cm$^{-1}$T$^{-1}$, it should be divided by 3298.0, yielding intensities around four orders of magnitude smaller than the computed ones, one of the reason of such disagreement can be normalization of experimental data to 1 Gauss.
% The experimental MCD spectrum (normalized to 1 Gauss), reported by~\citeauthor{910Anthra_exp2}~\cite{910Anthra_exp2} for 9,10-anthraquinone supports this assumption. Thus we have robust elements to the MCD experimental units in Ref.~\citenum{exp:Anthra:Naphta} were misprinted and that the scaling factor missing is exactly 10$^{4}$, this was taken into account in our discussion.}

The discussion of the results in the following section is based on spectra shifted in order to realign with experiment. Details on the shifts applied are given in the figure captions.
{Tables of raw data (transition energies, oscillator strengths and ${\mathcal{B}}$ terms) are reported in the SI file.}

%The experimental data for 2-thiouracil, taken from \citenum{hypo_exp}, was scaled with factor 10$^{3}$ following \citenum{thiour_exp}.

\section{Results and Discussion}
\label{Results}
% In this section, we discuss our ADC results for the MCD spectra of uracil, 2-thiouracil, 4-thiouracil, purine, hypoxanthine, \mbox{1,4-naphthoquinone} and  9,10-anthraquinone,
%
% The following bit is also a repetition
%
% In this section, we discuss our ADC results for the MCD spectra of the 
% uracil, 2-thiouracil, 4-thiouracil, purine, hypoxanthine, \mbox{1,4-naphthoquinone},  \mbox{9,10-anthraquinone}, and \mbox{1-naphthylamine} molecules,
% %(in gas phase). 
% %computed using our new MCD-ADC($n$) implementation, and 
% comparing them to RICC2 results as well as experimental data.
% %We also present OPA spectra for help with the interpretation of the MCD spectrum. 
% %As anticipated, a
% All spectra presented in the section have been shifted with respect to experiment for ease of comparison. Tables with data computed at ADC(2)-x level can be found in the ESI file.
% The unshifted spectra can be found in the ESI file. 

\subsection{Uracil, 2-thiouracil, 4-thiouracil}

Experimental OPA and MCD spectra of uracil and thiouracils in water were recorded by \citeauthor{OPA+MCD_exp};~\cite{OPA+MCD_exp} and \citeauthor{thiour_exp},~\cite{thiour_exp} respectively. 

A computational study of the MCD spectrum of uracil (and other molecules) was presented by \citeauthor{CPP:MCD:Nucleic}~\cite{CPP:MCD:Nucleic} at the TD-DFT level of theory. Results from simulations both \textit{in vacuo} and in water solution were reported, the latter described using the polarizable continuum model (PCM) with explicit water molecules.
%
% Experimental OPA and MCD spectra of {\bf uracil} were recorded in water by  \citeauthor{OPA+MCD_exp};~\cite{OPA+MCD_exp}
% those of the thiouracils, also recorded in water, are taken from \citeauthor{thiour_exp}~\cite{thiour_exp}
%
%\alert{I believe the MCD of uracil is not from  Igarashi...}
% The OPA spectrum of uracil in gas phase has been measured by \citeauthor{ur_exp_vapor}.~\cite{ur_exp_vapor} \revD{do we really need it? spectrum is normalized}
% A computational study of the MCD spectrum of uracil (and other molecules) was presented by \citeauthor{CPP:MCD:Nucleic}~\cite{CPP:MCD:Nucleic} at the TDDFT level of theory. Results were reported both \textit{in vacuo} and in water solution, the latter described using the polarizable continuum model (PCM) with explicit water molecules.
A detailed computational investigation of OPA and MCD spectra of various thiouracils, including 2-thiuracil {and 4-thiouracil}, was carried out by~\citeauthor{thiouracil_in_solution},~\cite{thiouracil_in_solution} also at the TD-DFT level \textit{in vacuo} and in PCM/water.

\begin{figure*}[hbt!]
\centering
\includegraphics[width=\linewidth]{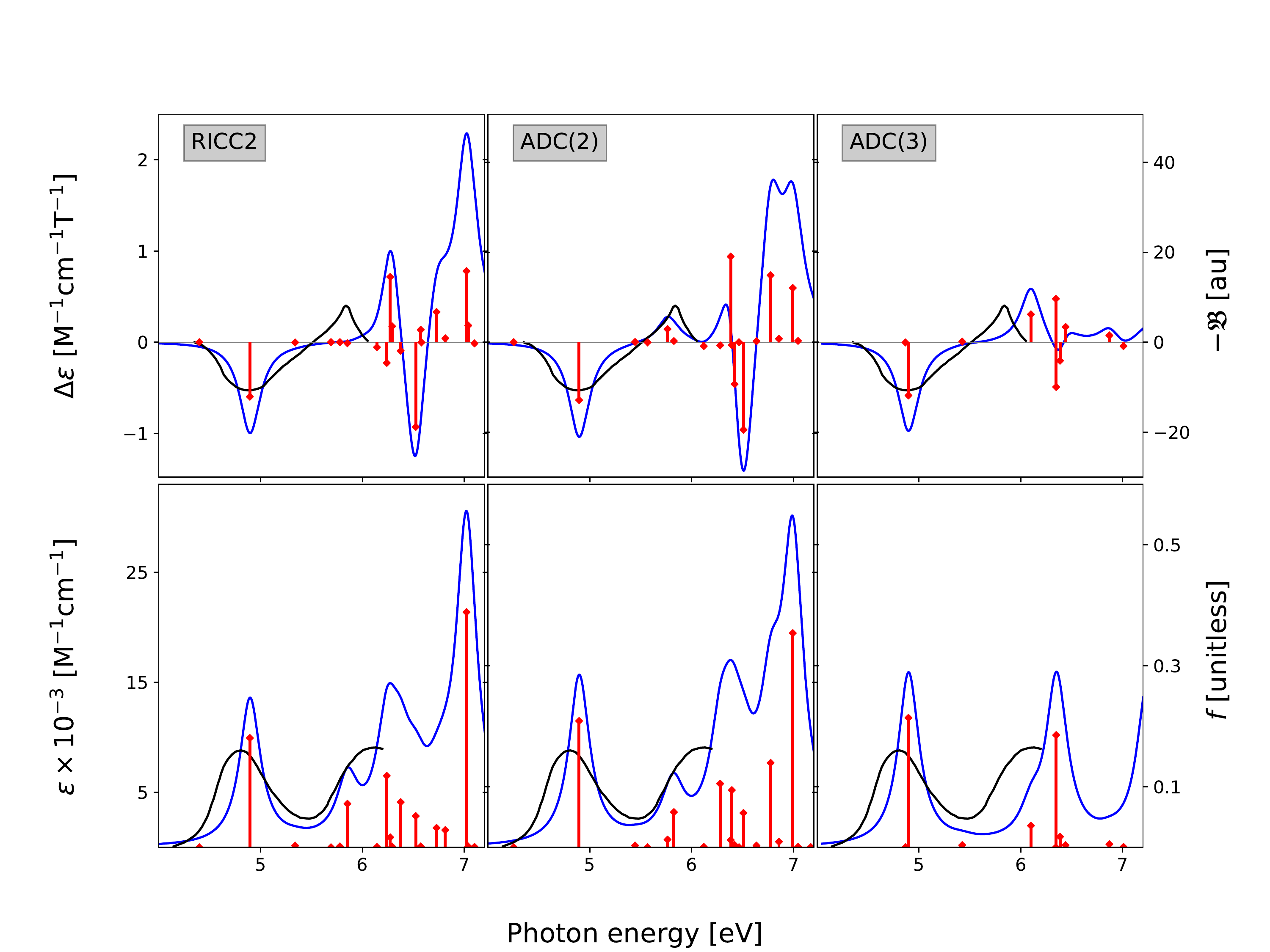}
\caption{Uracil. MCD (upper panels) and OPA spectra (lower panels) at RICC2, ADC(2) and ADC(3) levels. The experimental spectra recorded in water (black lines) were digitized from Ref.~\citenum{OPA+MCD_exp}. 
The simulated spectra were shifted to align with experiment by $-0.5$ eV (RICC2), $-0.4$ eV (ADC(2)) and $-0.45$ eV (ADC(3)).
% Both simulated spectra were rigidly shifted by +19.2 nm (ADC(2)) and +23.5 nm (CC2) along the x axis to align with experiment.
\label{fig:Ur:ADC2:CC2:shift}}
\end{figure*}

% \begin{figure}[hbt!]
% \centering
% \includegraphics[width=\linewidth]{Figures/Uracil_ADC2x_ADC3_shift.pdf}
% \caption{Uracil. MCD (upper panels) and OPA spectra (lower panels) at ADC(2)-x and ADC(3) level. Experimental spectra recorded in water, 
% %and in vapor~\cite{ur_exp_vapor}
% reported as black lines,
% were redigitized from Ref.~\citenum{OPA+MCD_exp}. The simulated spectra were shifted by $-$17.6 nm and +21.5~nm, for ADC(2)-x and ADC(3), respectively,
% to roughly align with the experiment. The first dark state predicted by ADC(2)-x at 318 nm is not shown. \label{fig:Ur_CC2_ADC2:shift}
% %+19.2 nm for ADC(2), $-$17.6 nm for ADC(2)-x, +21.5 nm for ADC(3) and +23.5 nm for CC2. \
% \label{fig:Ur:ADC2x:ADC3:shift}
% }
% \end{figure}

Fig.~\ref{fig:Ur:ADC2:CC2:shift}, %and~\ref{SI-fig:Ur:2thio:4thio:ADC2x}, 
top panels, collects the MCD spectra computed at RICC2, ADC(2) and ADC(3) levels of theory.  Fig.~\ref{SI-fig:Ur:2thio:4thio:ADC2x} shows the spectra  at the ADC(2)-x level. 
In the figures, we also report the experimental MCD spectrum of uracil (black line), recorded by \citeauthor{OPA+MCD_exp} in water~\cite{OPA+MCD_exp} in-between % 
4.5 eV (280 nm)
and 6.0 eV (205 nm).
%205 nm. 
This shows a negative band centered at around 
%255 nm
4.9 eV (255 nm)
and a positive one at around 5.8 eV (215 nm).
The bottom panels show the OPA spectra, {again with the black line as the experimental spectrum recorded in water,~\cite{OPA+MCD_exp} which shows one well-separated feature
centered at $\approx$4.8 eV (260 nm), 
plus the offset of a second peak.} 

\begin{figure}[hbt!]
\centering
\includegraphics[width=1.0\linewidth]{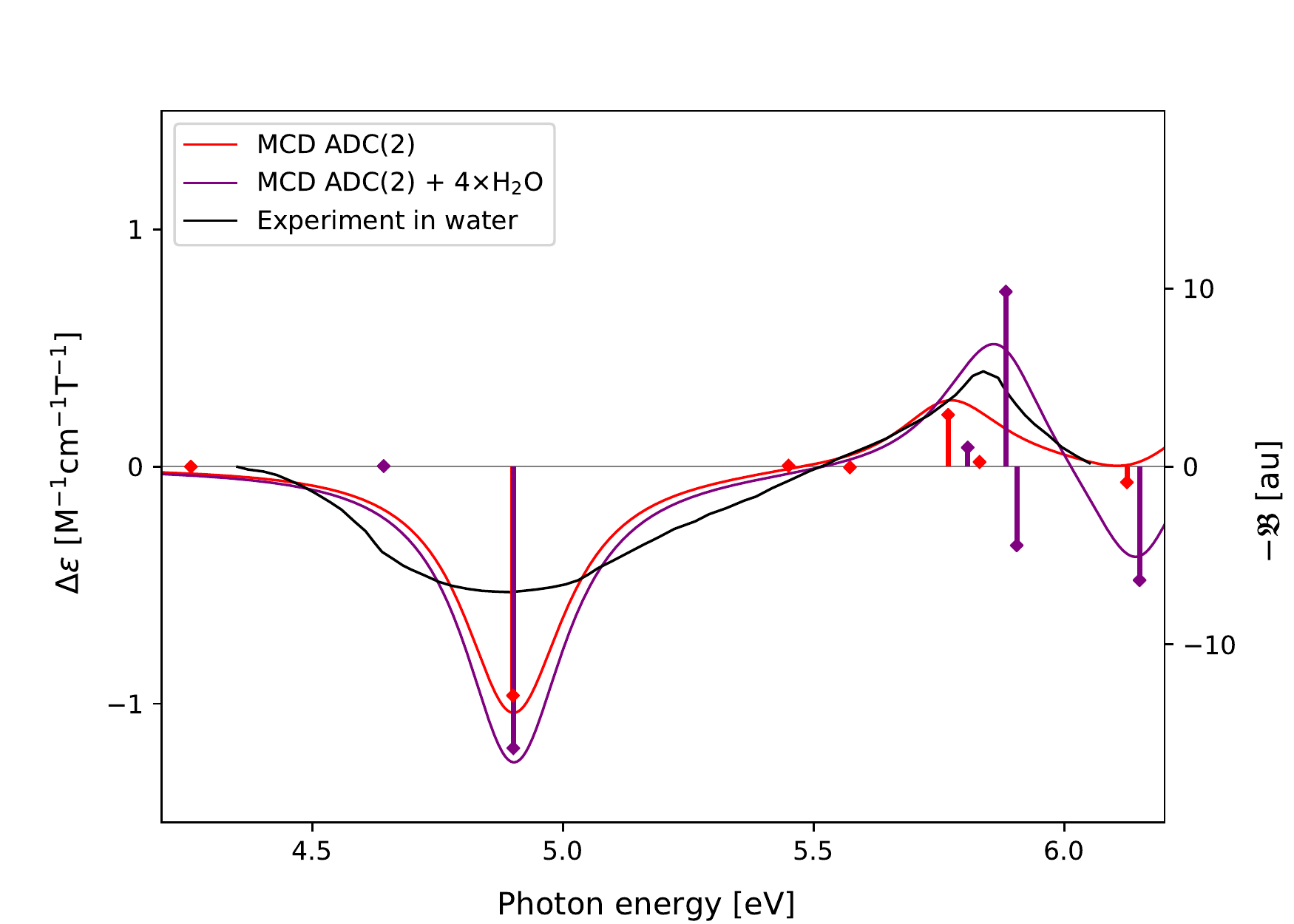}
\caption{Uracil. Comparison of MCD 
at ADC(2) level in gas phase and 
for a model system including 4 explicit water molecules. 
The geometry of the latter was taken 
from Ref.~\citenum{CPP:MCD:Nucleic}. The experimental spectrum recorded in water~\cite{OPA+MCD_exp} is also given. 
The computed spectra were shifted to align with the lowest energy experimental band, by 
%19.4 nm and 12.8 nm 
$-0.4$ eV and $-0.25$ eV
for gas phase and explicit solvation model, respectively.\label{fig:ur+w}
}
\end{figure}

%Two experimental OPA spectra, one recorded in water~\cite{OPA+MCD_exp} and one in vapour,~\cite{ur_exp_vapor} are reported. Both spectra show one well-separated feature plus the offset of a second peak.
% \revS{???I don't understand}% \revD{Here I would like to say, that both experimental spectra have similar profile and both spectra show the first feature and a half (only shoulder) of the second feature}
% \revS{it's not a shoulder, is the beginning of some other peak}\\
Keeping in mind that the experimental spectra were recorded in water, whereas the calculations were done without solvent model, all simulated MCD spectra roughly reproduce the shape of the experimental spectrum.
%general description
All methods attribute the first peak in both OPA and MCD to the bright $\pi\to\pi^*$ {(1A')} excited state, well separated from the other excitations. 
The origin of the second feature, however, varies significantly between the four methods. 
In particular, 
ADC(2) and ADC(3) attribute it primarily to a  transition to the 2A' state, 
whereas  in
RICC2 the positive band comes from the balance between {transitions of opposite signs}, with 4A' and 5A'' as the main contributors.
ADC(2)-x yields an almost negligible intensity of the second feature.
Remarkably, the ADC(2) spectrum presents two positive bands in the mid-region (in between
{5.5 eV--6.5 eV}
% \revS{I DON'T SEE TWO POSITIVE BANDS BETWEEN 5.5 AND 6!!!}
%200-220 nm 
in the shifted spectrum in Fig.~\ref{fig:Ur:ADC2:CC2:shift}). 
However, as it can be seen from Fig.~\ref{fig:ur+w}, when including
four molecules (corresponding to the first solvation shell), using the optimized configuration of Ref.~\citenum{CPP:MCD:Nucleic}, the additional peak disappears and the agreement with experiment improves. 

% Remarkable, the position of the first dark excited state, $n\to\pi^{\ast}$ (1A"), varies between methods, but all four methods predict zero MCD intensity and oscillator strength. As a result, $n\to\pi^{\ast}$ does not make any contribution to broaden OPA and MCD spectra. In this respect, this observations of the first dark and first bright transitions are in agreement with our previous study~\cite{Uracil:Daniil} where both states of uracil were intensively studied with various computational techniques.}  

In the experiment, the separation between the first and second band is $\sim$0.94 eV.
%$\sim$40 nm.
Among the four computational methods, RICC2 yields (\textit{in vacuo})
%four MCD features: the first well-separated negative peak at the beginning of the spectra and then positive, negative, and positive peaks. The method predicts about 1.5 times the higher intensity of the first spectral feature and twice as higher intensity of the second one,  with respect to recorded spectrum, with 
the largest separation, $\sim$1.4 eV  
%$\sim$50 nm,  
while ADC(2)-x predicts the smallest one, around 0.6 eV.
%30 nm. 
The ADC(2) method estimates this separation to be $\sim$0.87 eV 
%$\sim$35 nm, 
and ADC(3)  predicts a distance between the first and second feature of approximately 1.2 eV. 
%43 nm.
The ADC(2) calculation with four explicit solvent molecules yields a peak separation of 1 eV, 
which is the result closest to experiment.
%reproduces the two first features of the experimental spectrum, the remaining part of the spectrum is presented with two transitions of the same magnitude but different signs which cancel each other, and a number of positive low-intensity peaks at the end of the spectrum. This method shows the closest intensity of the second feature and 
%Similar to other methods, 
%reproduces two first features, but the rest part of the spectrum differs quite a lot  with respect to experiment. This method 
% underestimates the relative position between the first and second features and predicts the almost indistinguishable intensity of the second peak.
%should we say a couple of words about OPA?
Thus, explicit solvation
slightly improves the relative distance between first and second peak  of uracil, at least at ADC(2) level.
Our attempts to cover a larger region 
were not successful due to convergence problems in the response equations for a higher number of roots.
%, as already discussed in previous studies.~\cite{ADC3compareFCICC,ADC3:1999,ADC3:2014}

Comparing to the 
TD-DFT~\cite{CPP:MCD:Nucleic} results \textit{in vacuo} and in water solution reported in 
Ref.~\citenum{CPP:MCD:Nucleic},
the TD-CAM-B3LYP~\cite{CPP:MCD:Nucleic} spectrum is in agreement with all the ADC spectra, as well as the RICC2 ones. TD-B3LYP, on the other hand, shows a second negative feature at around 6.2 eV 
%200 nm, 
not present in the experimental spectrum (in solution). For both ADC(2) and TD-DFT,~\cite{CPP:MCD:Nucleic}explicit consideration of a few water molecules improves the agreement with experiment.
% Going towards higher energies, the differences between the MCD spectra by the four methods increases considerably. 
%\citeauthor{CPP:MCD:Nucleic}\cite{CPP:MCD:Nucleic} reported  computed MCD spectrum at TD-B3LYP and TD-CAM-B3LYP levels in gas phase and estimation of solvent effect (PCM and explicit solvation model). 
% Similar to TD-DFT~\cite{CPP:MCD:Nucleic} results, both gas-phase and water solution calculation overestimates the intensities of both experimental features. In gas-phase, the
% \alert{I also suggest to move extended variant of ADC(2) in SI file, since it usually far from experimental picture, except first features. }

\begin{figure*}[hbt!]
\centering
\includegraphics[width=\linewidth]{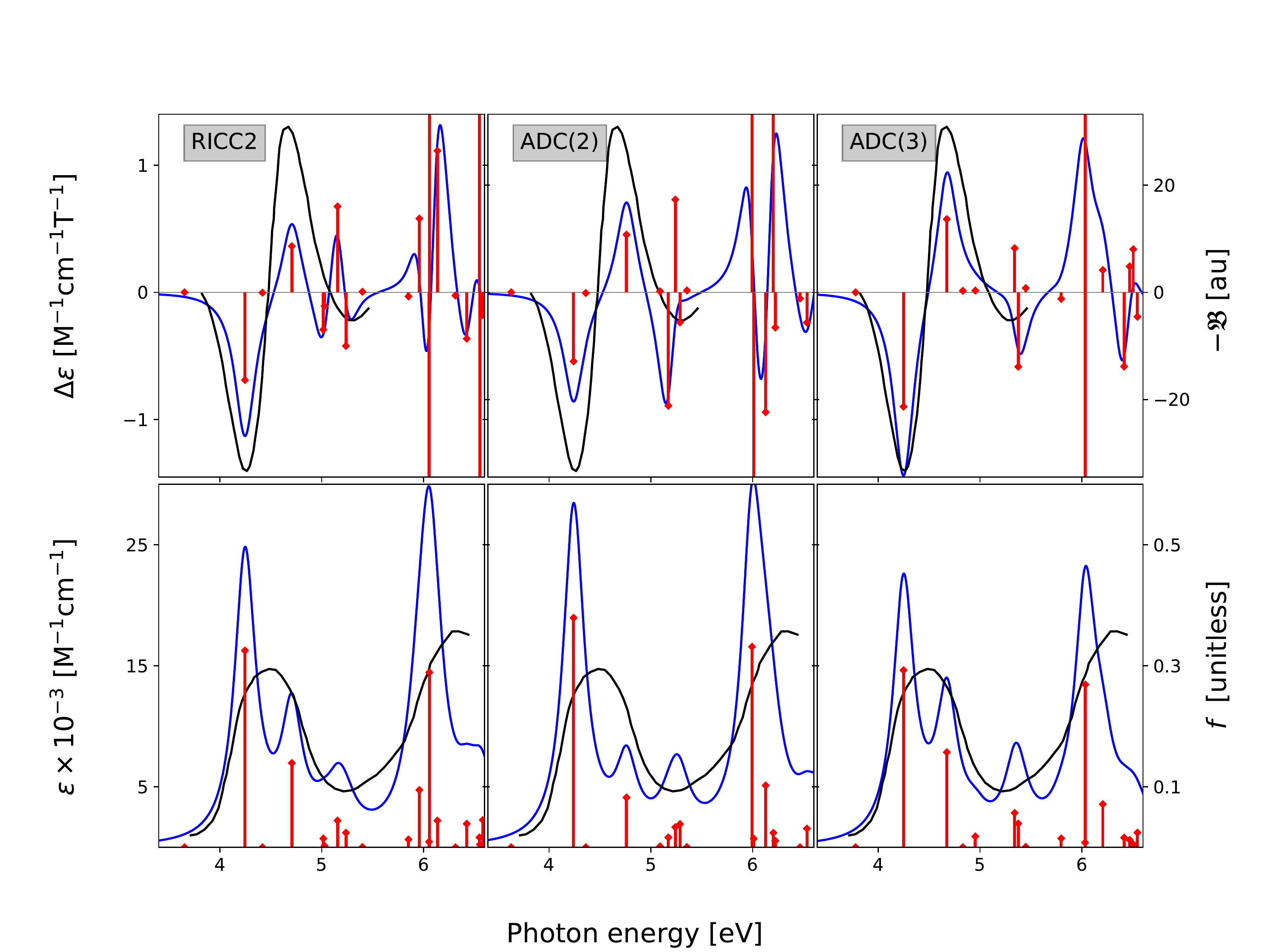}
\caption{2-thiouracil. MCD  (upper panels) and OPA spectra (lower panels) at RICC2, ADC(2) and ADC(3) levels. 
{The experimental spectra recorded in water, shown as black lines, were digitized from Ref.\citenum{thiour_exp} and rescaled
%scaled with a factor 10$^{3}$ in case of MCD, %and with a factor 10$^{1}$ for OPA, 
following Ref.~\citenum{thiouracil_in_solution}.}
The simulated spectra were shifted to align with experiment, by $-0.38$ eV (RICC2), $-0.23$ eV (ADC(2)) and $-0.38$ eV (ADC(3)).
% The simulated spectra were shifted by +14.3 nm (ADC(2)) and +23.6 nm (CC2) to roughly align with experiment. 
%\revS{I DO NOT UNDERSTAND THESE SHIFTS} 
\label{fig:2thioUr:ADC2_CC2:shift}
%\revD{Peaks positions according to table in exp article are 290 ($-$0.46), 263 (0.43), 237 ($-$0.07) so figure is fine.}
}
\end{figure*}

%2-thiouracil
% Fig.~\ref{fig:2thioUr:ADC2_CC2:shift} 
% %and~\ref{fig:2thioUr:ADC2x_ADC3:shift}
% show the computed OPA and MCD spectra of \textbf{2-thiouracil}. 
Fig.~\ref{fig:2thioUr:ADC2_CC2:shift} 
 %and~\ref{fig:2thioUr:ADC2x_ADC3:shift}
 show the computed OPA and MCD spectra of \mbox{\textbf{2-thiouracil}}.
The experimental MCD spectrum, recorded in water,~\cite{thiour_exp} covers the region from $\sim$3.8 eV (325 nm) 
to $\sim$5.5 eV (225 nm) 
and shows a very distinctive bisignate feature, with negative and positive lobes peaking at 
$\sim$4.3~eV ({290} nm) 
and $\sim$4.7~eV,
({263} nm), 
respectively.
A third (negative) band of lower intensity is peaked at 5.2~eV (237 nm).
% \revS{recheck these values match on the plot!}. 
Due to molecular symmetry, the bisignate MCD band is not attributable to an $\mathcal{A}$ term.
It  
%clearly 
corresponds to a broad peak centered at around 
4.5 eV
%274 nm 
in the experimental OPA spectrum,~\cite{thiour_exp} revealing that more than one excited state contributes to the OPA band. 
The third, low intensity, MCD peak also appears to correspond to a transition 'hidden' between the first broad OPA band and the second OPA peak emerging at around 6.2 eV.
%200 nm 
{In the experimental study, the MCD spectrum was interpreted based on the hypothesis that a tautomeric equilibrium exists between thiol and thione forms of 2-thiouracil in solution. Thus, the band at 290 nm was tentatively assigned to a thiol form, and the one at $\approx$270 nm to a thione form. 
Another possibility to elucidate the spectrum mentioned in the experimental study is that the two peaks are due to two ``local'' 
$\pi\pi^*$ transitions, one on the thiourea moiety and one on the acrolein moiety. This interpretation was disregarded as too difficult to verify experimentally.}

All four methods reproduce the bisignate feature observed in the experimental spectrum, and confirm the presence of two electronic excitations within the first OPA band. Also, all methods indicate the presence of an additional negative band moving towards higher energies, which contains contributions from several excited states. 
Inspection of the natural transition orbitals (NTOs) for the two dominant excitations (see {Figure~\ref{SI-fig:2-thio:NTOs})} reveals that the first (negative) MCD band is a $\pi\pi^*$ where $\pi^*$ involves the S atom, whereas the transition responsible for the positive 
band is a $\pi\pi^*$ one, where $\pi^*$ involves the C=O group. This seems to support the second interpretation of the origin of the bisignate feature.

As observed for uracil, the high energy region of the spectra shows remarkably different MCD patterns between the four methods, and we refrain from a detailed discussion, also because the experimentally recorded spectra do not extend in such region.

% \begin{figure}[hbt!]
% \centering
% \includegraphics[width=\linewidth]{Figures/2thiour_adc2x_ADC3_shift.pdf}
% \caption{2-thiouracil. MCD (upper panels) and OPA spectra (lower panels) at ADC(2)-x and ADC(3) level. Experimental spectra recorded in water, shown as black lines, were re-digitized from Ref.\citenum{thiour_exp} and scaled
% %scaled with a factor 10$^{3}$ in case of MCD, %and with a factor 10$^{1}$ for OPA, 
% as in Ref.~\citenum{thiouracil_in_solution}.
% The simulated spectra were shifted by $-$39.5 nm (ADC(2)-x) and +23.7 nm (ADC(3)) 
% to roughly align with experiment.
% \label{fig:2thioUr:ADC2x_ADC3:shift}}
% \end{figure}

{Similar to the
TD-DFT~\cite{thiouracil_in_solution} results,
%for 2-thiouracil and 4-thiouracil 
in gas-phase the computed intensities are smaller than the experimental ones, except for those yielded by the ADC(3) method. ADC(3) provides an almost identical intensity for the first band and a small underestimation of the second one.}

\begin{figure*}[hbt!]
\centering
\includegraphics[width=\linewidth]{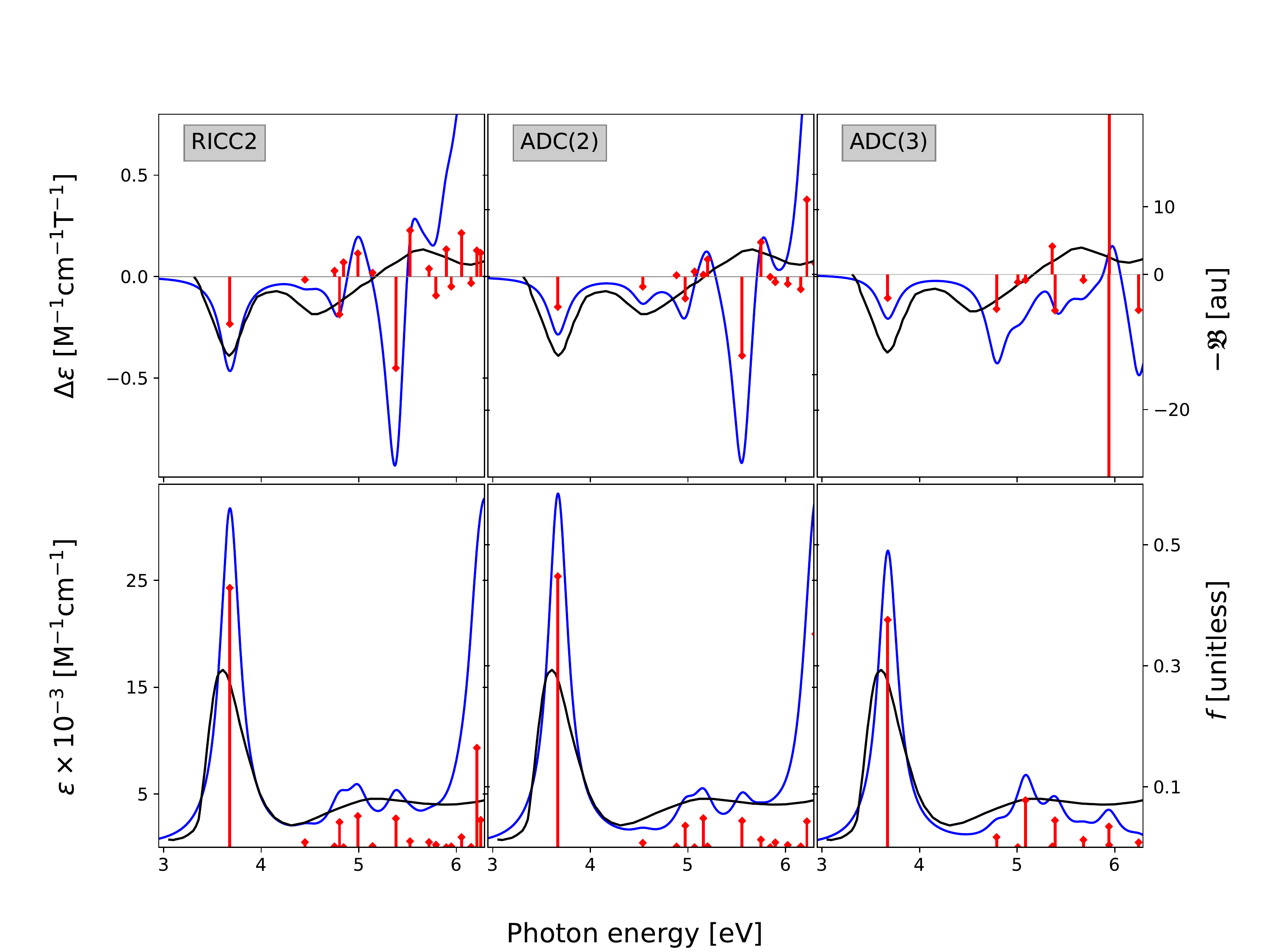}
\caption{4-thiouracil. MCD  (upper panels) and OPA spectra (lower panels) for RICC2, ADC(2) and ADC(3). Experimental spectra recorded in water are presented as black solid lines. The simulated spectra were shifted to align with experiment by $-0.6$ eV (RICC2), $-0.38$ eV (ADC(2)) and $-0.34$ eV (ADC(3)).
{The experimental spectra were digitized from Ref.\citenum{thiour_exp} and rescaled as in Ref.~\citenum{thiouracil_in_solution}.
% The experimental spectra, re-digitized from Ref.\citenum{thiour_exp} were scaled with a factor 10$^{3}$ for MCD and with a factor 10$^{1}$ for OPA, following Ref.~\citenum{thiouracil_in_solution}.
All three methods also yield a low-energy dark transition, which (once shifted) falls outside the frequency range in the figures.
% \revD{Note that optically dark transitions at 2.46 eV, %3.08-0.6 eV
% 2.49 eV
% %2.87-0.38 eV
% and 2.73 eV (after corresponding shift)
% %3.07-0.34 eV 
% predicted by RICC2, ADC(2), ADC(3), respectively, fall outside the figures range.
\label{fig:4thioUr:ADC2_CC2:shift}}}
\end{figure*}

The results at the RICC2, ADC(2) and ADC(3) levels  of theory for \textbf{\mbox{4-thiouracil}} are shown in Fig.~\ref{fig:4thioUr:ADC2_CC2:shift}. 
The \mbox{ADC(2)-x} spectra are in the SI.
%and~\ref{fig:4thioUr:ADC2x_ADC3:shift}. 
The spectral profiles are  distinctively different from those of \mbox{2-thiouracil}, 
highlighting the ability of MCD spectroscopy to discriminate between the two thiouracil isomers, as well as from uracil. The lower-energy region of the MCD spectrum of \mbox{4-thiouracil} presents 
two negative bands {($\sim$3.8 eV and $\sim$4.8)}
%{($\sim$320 nm and $\sim$260 nm)}, 
and a positive one at around 5.5 eV.
%two positive ones emerge going below 240 nm. 
The overall intensity of the MCD spectrum is, according to the experiment,~\cite{thiour_exp} weaker than in 2-thiouracil.
Both trends are reproduced in our calculations, even though for ADC(2)-x and ADC(3) the second negative band is as intense or even more intense than the first band.
One should keep in mind, however, that the experiment was recorded in water, so the comparison between theory and experiment can only be qualitative. For all four methods, the first negative band is attributed to one, well-separated, electronic transition, whereas several transitions may be contributing to the second band. 

Moving towards higher energies, the experimental and simulated spectra deviate from each other. As before, we do not discuss this region, as solvent effects are 
not accounted for in the computed spectra.

% \revS{We need to say a few words on the comparison with former DFT results (we mention  in the abstract that we will do so!)}

%(\revD{})  
% \begin{figure}[hbt!]
% \centering
% \includegraphics[width=\linewidth]{Figures/4thiour_adc2x_ADC3_shift.pdf}
% \caption{
% 4-thiouracil. MCD (upper panels) and OPA spectra (lower panels) for ADC(2)-x and ADC(3) electronic structure methods. Experimental spectra recorded in water are presented as black solid lines. Simulated spectra were shifted to align with experiment: $-$67 nm and +12 nm for ADC(2)-x and ADC(3), respectively. 
% {The experimental spectra were re-digitized from Ref.\citenum{thiour_exp} and modified 
% %were scaled with a factor 10$^{3}$ and with a factor 10$^{1}$ for OPA, 
% following Ref.~\citenum{thiouracil_in_solution}.} %\revS{WHAT DOES THIS MEAN?}
% % \revD{The dark transitions located at 598.7 nm and 403.6 nm for ADC(2)-x and ADC(3) methods, respectively, are out of the figure.
% {Note that an optically dark transition, located at 598.7 nm (ADC(2)-x) and 403.6 (ADC(3)), falls outside the figure's wavelength range.} \label{fig:4thioUr:ADC2x_ADC3:shift} }
% \end{figure}

%\clearpage

\subsection{Purine and hypoxanthine}
\begin{figure*}[hbt!]
\centering
\includegraphics[width=\linewidth]{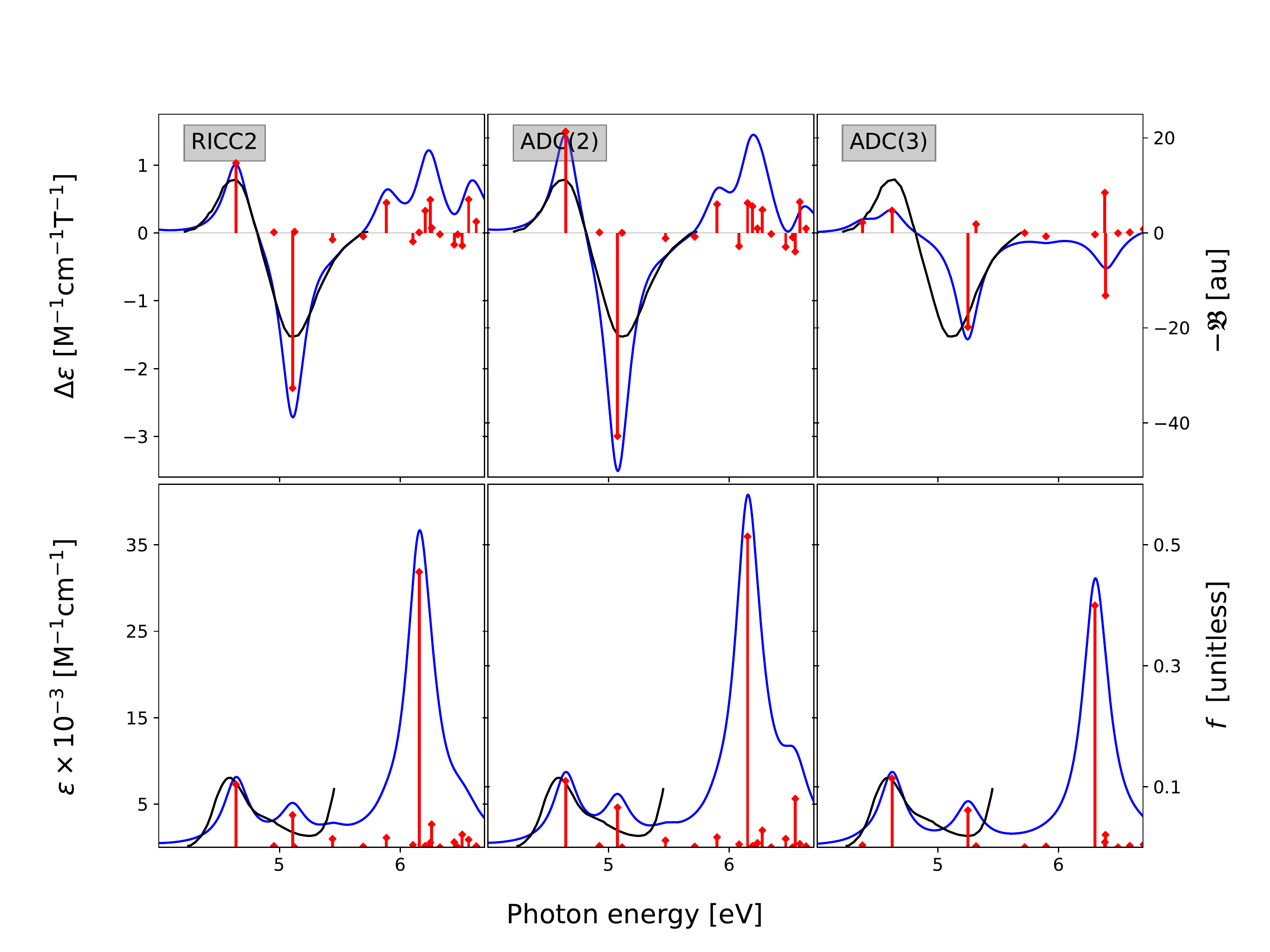}
\caption{Purine. MCD  (upper panels) and OPA  (lower panels) at RICC2 and ADC(2) level. Experimental spectra recorded in water are presented as black lines, and were digitized from Ref.\citenum{Pur_exp}. The simulated spectra were shifted to align with experiment by 
%+28.58 nm and +27.5 nm for RICC2 and ADC(2), respectively.
$-0.54$ eV (RICC2), $-0.51$ eV  (ADC(2)) and $-0.34$ eV (ADC(3)).
We note that ADC(2) and RICC2 also predict a low-energy optically dark transition (located at $\sim$3.9 eV after shifting), which falls outside the frequency range plotted.\label{fig:Pur:ADC2:CC2:ADC3:shift}}
\end{figure*}

\begin{figure*}[hbt!]
\centering
\includegraphics[width=\linewidth]{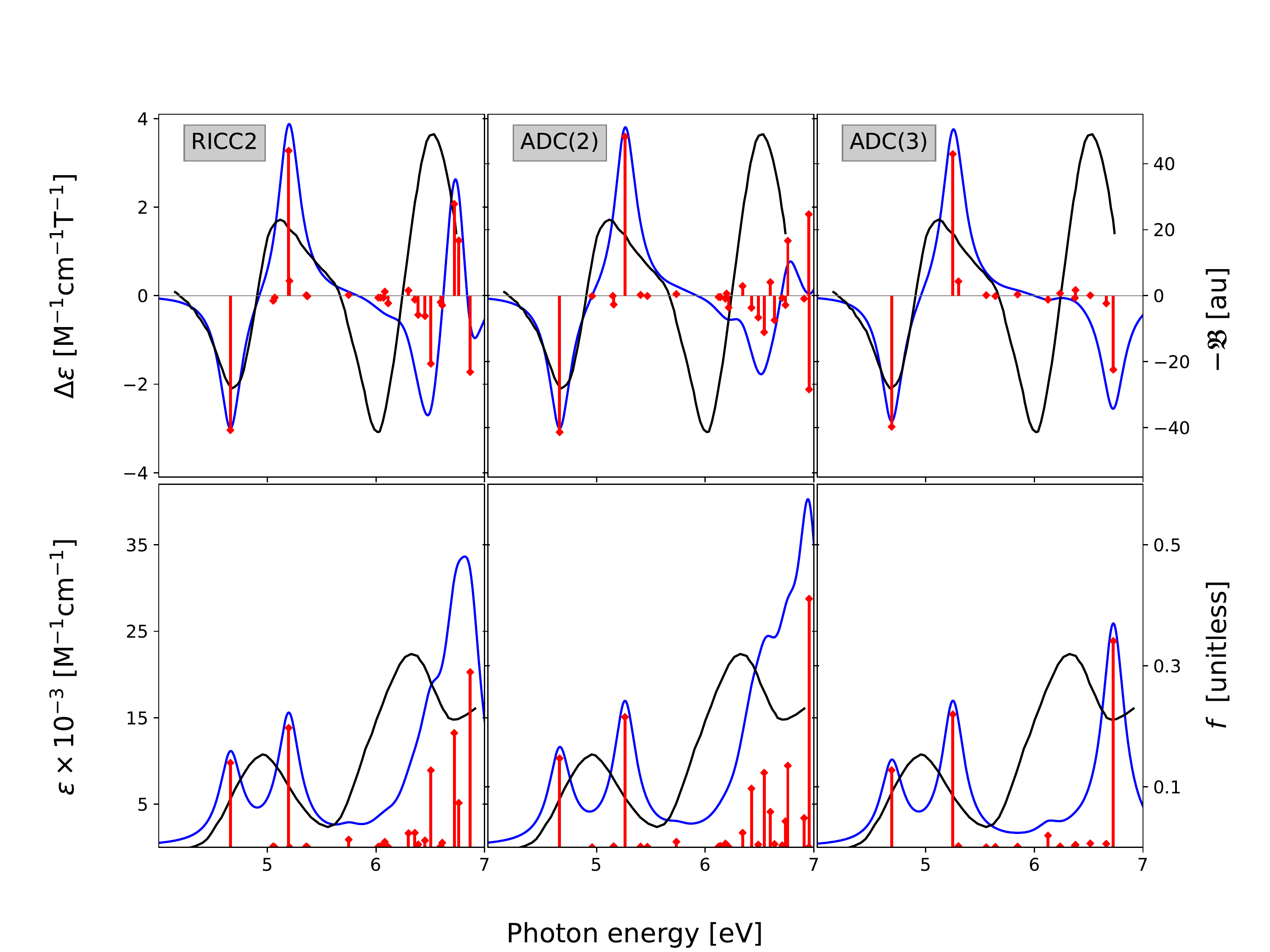}
\caption{Hypoxanthine. MCD (upper panels) and OPA (lower panels) at RICC2 and ADC(2) levels. The experimental spectra recorded in water (black lines) were digitized from Ref.\citenum{hypo_exp}. 
The simulated spectra were shifted to align with experiment: --0.39 eV, --0.26 eV and --0.22 eV for RICC2, ADC(2) and ADC(3) respectively.
% The simulated spectra were shifted by +19.2 nm (RICC2) and +13.1 nm (ADC(2)) to roughly align with experiment. 
\label{fig:Hypo:ADC2:CC2:shift}}
\end{figure*}

The OPA and MCD spectra of \textbf{purine} were recorded by \citeauthor{Pur_exp}\cite{Pur_exp} in water.
A computational investigation at the TD-DFT level both \textit{in vacuo} and using explicit and implicit solvation models was presented in Ref.~\citenum{CPP:MCD:Nucleic}. Purine was also studied at the CC2 and CCSD level \textit{in vacuo} and using COSMO in Ref.~\citenum{Sarah:MCD}.
Experimental spectra data of the purine derivative hypoxanthine have been presented in water by \citeauthor{hypo_exp}\cite{hypo_exp} and by \citeauthor{OPA+MCD_exp}
\cite{OPA+MCD_exp}
Hypoxanthine was the subject of computation analysis in Ref.~\citenum{CPP:PE:MCD} at the Polarizable-Embedding TD-DFT level, and also at TD-DFT level using the polarizable continuum model (PCM) with and without explicit water molecules in Ref.~\citenum{CPP:MCD:Nucleic}.
Our computed spectra 
at RICC2, ADC(2) and ADC(3) level \textit{in vacuo} 
for both molecules are presented in Figs.~\ref{fig:Pur:ADC2:CC2:ADC3:shift} and~\ref{fig:Hypo:ADC2:CC2:shift}.
{Fig.~\ref{SI-fig:Pur:Hypo:ADC2-x}} shows the ADC(2)-x spectra. 

% \begin{figure}[hbt!]
% \centering
% \includegraphics[width=\linewidth]{Figures/Pur_ADC2x_ADC3_shift.pdf}
% \caption{Purine. MCD  (upper panels) and OPA   (lower panels) at ADC(2)-x and ADC(3) level. Experimental spectra recorded in water are presented as black lines,
% and were re-digitized from Ref.~\citenum{Pur_exp}. The simulated spectra were shifted, by $-$23.73 nm and +18.19 nm for ADC(2)-x and ADC(3), respectively, to roughly align with experiment.
% {An optically dark transition, located at 323.81 nm 
% %\revS{PLEASE WRITE THE VALUE!!} 
% (ADC(2)-x), falls outside the figure's wavelength range.}
% \label{fig:Pur:ADC2x:ADC3:shift}}
% \end{figure}

The experimental MCD spectrum of purine is characterized by a bisignate feature with the positive lobe in the low energy region. Both RICC2 and ADC(2) attribute this lowest energy peak
to a single excited state state {(1A')} with negative $\mathcal{B}$ term, whereas the negative lobe is due to the {fourth} 
%third 
excited state {(2A')} 
having a positive $\mathcal{B}$ term (For both methods, the first, optically dark state, falls outside the figures range). 
For ADC(3), both the first (1A'') and the second (1A') 
electronic excited states contribute to the first MCD band, even though the first excitation is practically dark in OPA. 
%\revD{Remarkable, that ADC(3) method predicts the larger intensity of the 1A" state, of compare  to other methods}
In ADC(2)-x the {fifth}
%fourth 
excited state is responsible for the negative MCD lobe (also in this case there is a first optically dark state outside the figures range). 
The larger intensity of the second peak over the first one is reproduced by all four methods.
{In the high energy region, RICC2 and ADC(2) predict several positive features, due to  multiple transitions, while ADC(2)-x predicts only one. In contrast to the other methods, ADC(3) predicts a negatively signed third band. Note that the experimental spectrum does not reach this region.}

% \begin{figure}[hbt!]
% \centering
% \includegraphics[width=\linewidth]{Figures/Purine+w_shift_1peakpo.pdf}
% \caption{Purine.
% Comparison of MCD spectra at ADC(2) level in vacuo and for a model system including 
% %\revS{5}
% \revD{4}
% explicit water molecules.
% The geometry of the latter was taken from Ref.~\citenum{CPP:MCD:Nucleic}.
% The experimental spectrum was re-digitized from Ref.\citenum{Pur_exp}.
% % \revS{PLEASE ALIGN TO THE INFLECTION POINT AND INDICATE THE SHIFTS IN THE CAPTION.} 
% \revD{The simulated spectra were shifted with respect to center of the first peak, by 28.0 nm and 28.0 nm for gas phase and explicit solvation model respectively, to roughly align with experiment.}
% \label{fig:Pur:CC2:ADC2:+w:shift_1centerpo}}
% \end{figure}

\begin{figure}[hbt!]
\centering
\includegraphics[width=\linewidth]{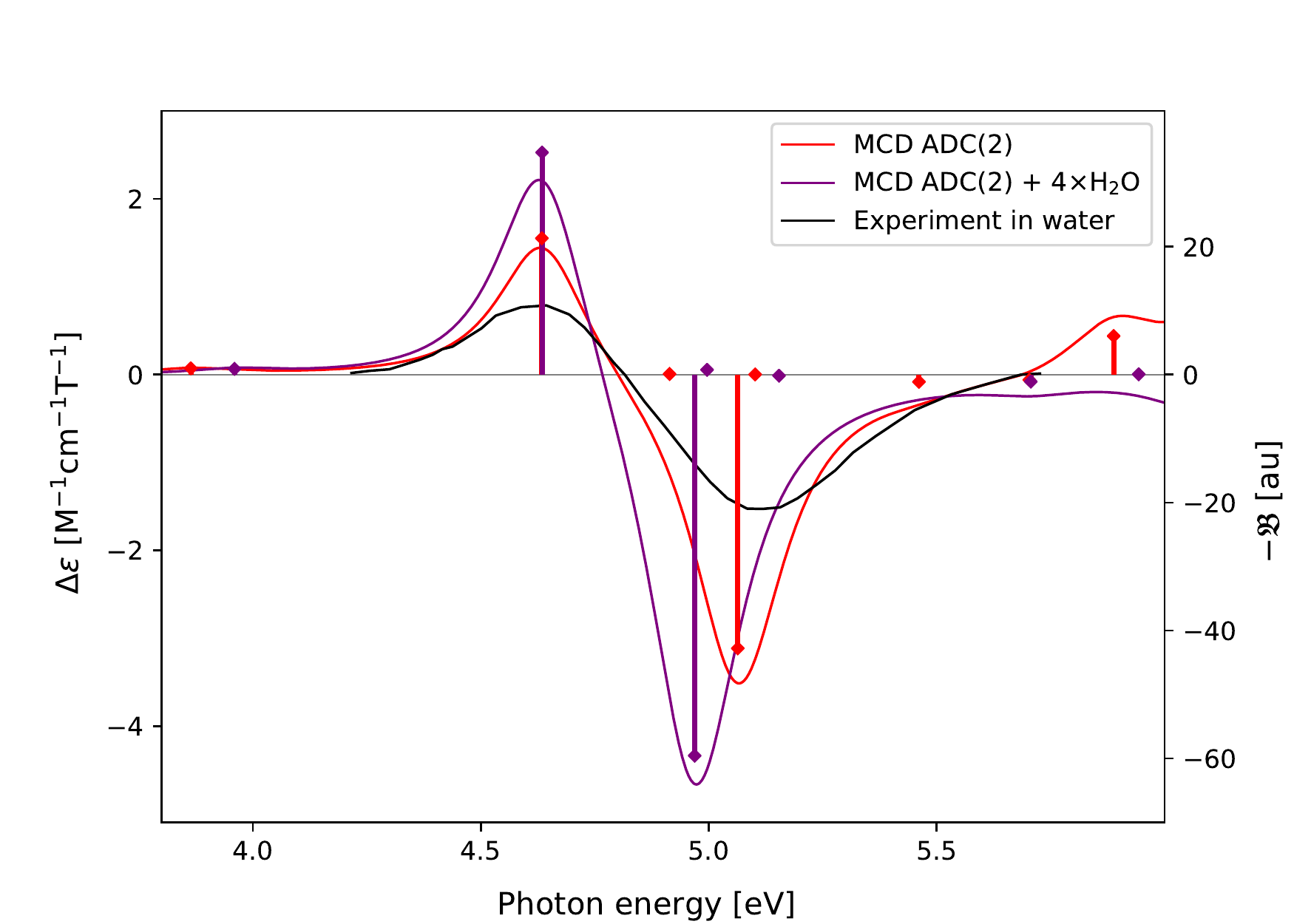}
\caption{Purine.
Comparison of MCD spectra at ADC(2) level \textit{in vacuo} and for a model system including 
4 explicit water molecules.
The geometry of the latter was taken from Ref.~\citenum{CPP:MCD:Nucleic}.
The experimental spectrum was digitized from Ref.\citenum{Pur_exp}.
{The simulated spectra were shifted by $-0.5$ eV (gas phase) and $-0.56$ eV (explicit solvation model) to align them to the inflection point in the experiment.} \label{fig:Pur:CC2:ADC2:+w:shift_infpo}}
\end{figure}

{For purine, the importance of solvent effects is illustrated, at ADC(2) level, in Fig.~\ref{fig:Pur:CC2:ADC2:+w:shift_infpo}.}
{Including the first solvation shell in the calculations increases the intensity of both features by approximately a factor of two and reduces the separation between the two bands, deteriorating the agreement with the experiment. 
We also refer to the study by  \citeauthor{Sarah:MCD}\cite{Sarah:MCD} for a discussion of solvent effects on the MCD spectrum of purine via COSMO.}

The spectra of hypoxanthine are presented in Fig.~\ref{fig:Hypo:ADC2:CC2:shift}.
%and \ref{fig:Hypo:ADC2x:ADC3:shift}.
The experimental MCD spectrum (once again in water) shows two bisignate features in-between %180-280 nm.
4.4--6.9 eV.
All methods reproduce the first ``pseudo-{$\mathcal{A}$}'' band, due to the 1A' and 2A' states; the ADC methods struggle with the  one at higher energy, in particular to reproduce the positive lobe. 
{According to RICC2, the fourth spectral feature is due to two close-lying transitions (11A' and 12A'').} 
In contrast, ADC(2) predicts two low-intensity positive peaks, formed by many transitions with 
{$\mathcal{B}$} terms of different sign and intensity.
As we are reaching higher energies, where many transitions contribute to the OPA and MCD bands, the stick-spectrum approach may become inefficient, due to the increasing number of excited states to be converged.
In support of our statement, we note that in the TD-DFT study reported by \citeauthor{CPP:MCD:Nucleic}\cite{CPP:MCD:Nucleic}
using the complex-polarization-propagator approach, where the MCD signal is directly computed, the second bisignate feature is indeed well reproduced.

%
% SONIA: DEAR DANIIL, DISCUSSING INTENSITIES 'VISUALLY' IS ALWAYS A BIT TRICKY, AS THEY DEPEND ON THE GAMMA PARAMETER YOU USE.
%
% \revD{The TDDFT study,  of both molecules shows similar results with ADC and RICC2 methods. All simulated spectra exhibit stronger intensities with respect to recorded spectra. The only is ADC(3) spectrum for purine. ADC(3) method shows a slight underestimation of the first feature and the good agreement in intensity for the second one. }

\begin{figure}[hbt!]
\centering
 \includegraphics[width=\linewidth]{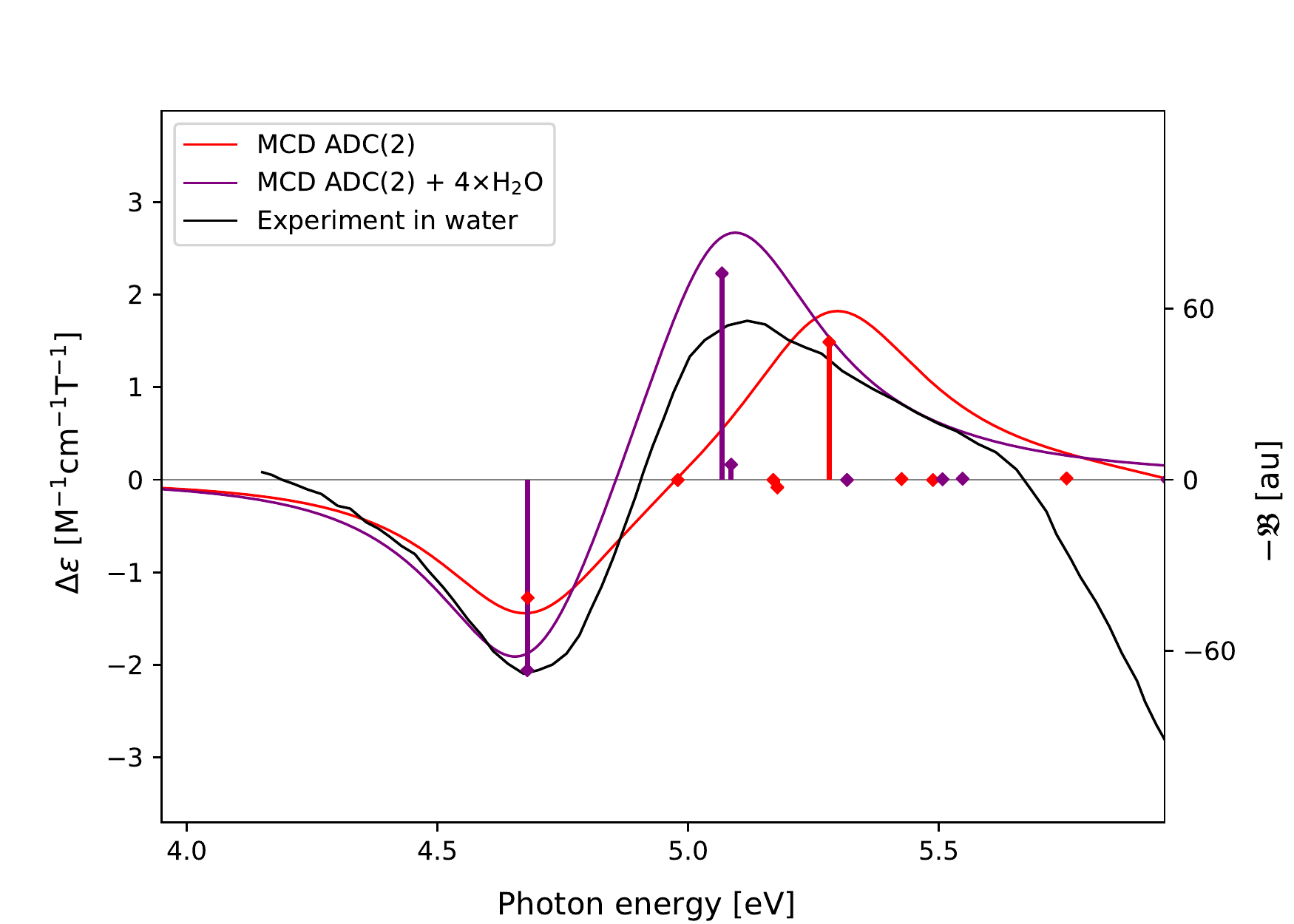}
\caption{
Hypoxanthine. Comparison of MCD spectra at ADC(2) level \textit{in vacuo} and for a model system including {5} explicit water molecules,
whose geometry was taken from Ref.~\citenum{CPP:MCD:Nucleic}. The experimental spectrum was digitized from Ref.\citenum{hypo_exp}. The simulated spectra were shifted to align them with the experimental inflection point at lowest energy,
by $-$0.24 eV
%14.5 nm 
(gas) and $-$0.28 eV 
%11.7 nm 
(solvated), respectively. The $\gamma$ parameter (see Eq.~\ref{broadening}) for this figure is 8.56 $\times$10$^{-3}$~a.u.\label{fig:Hypo:CC2:ADC2:+w:infpo_diffgamma}
}
\end{figure}

{The ADC(2) simulation using the explicit solvation model for hypoxanthine (Fig.~\ref{fig:Hypo:CC2:ADC2:+w:infpo_diffgamma}) reproduces the first and the second spectral features with slightly larger intensities. As in the analogous calculation for purine, the separation between the bright states becomes smaller, which, in {this specific case}, does improve a bit the agreement with the experiment.}
% {Note that the calculations including explicit water molecules only cover the first 50 nm, i.e., the region 210-260 nm after the realigning shift.}
% {Note that the calculations including explicit water molecules only cover 2 eV, i.e., the region 4-6 eV after the realigning shift.}

\subsection{1,4-naphthoquinone, 9,10-anthraquinone and {1-naphthylamine}}

As final examples, we considered two quinones, the smaller 1,4-naphthoquinone {(NQ)} and the larger 9,10-anthraquinone {(AQ)}, {and 1-naphthylamine (1NA)}. Experimental MCD spectra for NQ and AQ were reported, together with those for other ketones, by \citeauthor{exp:Anthra:Naphta}\cite{exp:Anthra:Naphta} and \citeauthor{910Anthra_exp2}.\cite{910Anthra_exp2} 
These results have earlier provided a reference for theoretical investigations at different levels of theory.\cite{MCD_Coriani_2000,exp:Anthra:Naphta,fukuda_14Naph_MCD,910Anthra_exp2,MCD:Xiaosong:2019,Seth2008:b}  
% For this work, NQ and AQ represent a particularly case study since a relatively recent study on both molecules at
A TD-DFT study,\cite{Seth2008:b} in particular, showed contradictory results with respect to the experimental data, which makes NQ and AQ interesting test 
cases for our MCD-ADC scheme.
Given the size of NQ and AQ, only RICC2 and ADC(2) are considered.

%NQ
{As in Ref.~\citenum{Seth2008:b}, we do not discuss the lowest energy transitions in NQ, which 
\citeauthor{exp:Anthra:Naphta}\cite{exp:Anthra:Naphta} attributed to n$\pi^*$.}
% A weak pseudo-{$\mathcal{A}$} feature was then found experimentally with its inflection point at around 320 nm.
Simulations of the MCD spectrum for NQ were reported by~\citeauthor{Seth2008:b}~\cite{Seth2008:b} 
at TD-DFT level,
and by \citeauthor{MCD:Xiaosong:2019}~\cite{MCD:Xiaosong:2019} using real-time (RT-) TD-DFT (B3LYP functional) 
and gauge-including atomic orbitals (6-31G(d) basis set).
The experimental spectrum of NQ includes three well-distinguishable features: a low-intensity pseudo-{$\mathcal{A}$} feature with inflection point at $\sim$3.8 eV,
%\sim$320 nm, 
a positive peak at around 4.9 eV,
%250 nm, 
followed by a series of  well-defined negative peaks of progressively increasing intensities. 
Both ADC(2) and RICC2 reproduce the two first experimental features, though with higher intensities. 
%More then that both methods predict %pseudo-{$\mathcal{A}$} to be formed with 1A1 and 1B1 states,
% 3B1 and 4A1 form two negative features  
%In both methods three features are formed by the same transitions. 
In correspondence to the positive band at around 5.1 eV, 
%250 nm, 
RICC2 shows a numerical instability, as two degenerate transitions with exceedingly large and oppositely signed ${\mathcal{B}}$ terms are obtained. Closer analysis reveals that the RICC2 results for the excitation energies of \mbox{1,4-naphthoquinone} are quite sensitive to the geometry used in the calculations.
In the high energy region, ADC(2) and RICC2 both predict  a strong positive peak (due to the 6B$_1$ transition) at around 6.5 eV,
%240-190 nm,
not present in the experimental spectrum (recorded in cyclohexane). 
{The spectra available at TD-DFT level are also in qualitative agreement with the experimental spectral profile of NQ. However, 
in the spectrum reported by~\citeauthor{MCD:Xiaosong:2019}~\cite{MCD:Xiaosong:2019} the relative intensity of the negative peaks is off compared to experiment. The SAOP/QZ3P1D and BP86/DZP spectra reported by~\citeauthor{Seth2008:b},~\cite{Seth2008:b} on the other hand, present a much better agreement of the relative intensities of these two negative bands. 
BP86/DZP fails in yielding the negative lobe of the low-energy
bisignate band of NQ.
}

\begin{figure*}[hbt!]
\centering
\includegraphics[width=\linewidth]{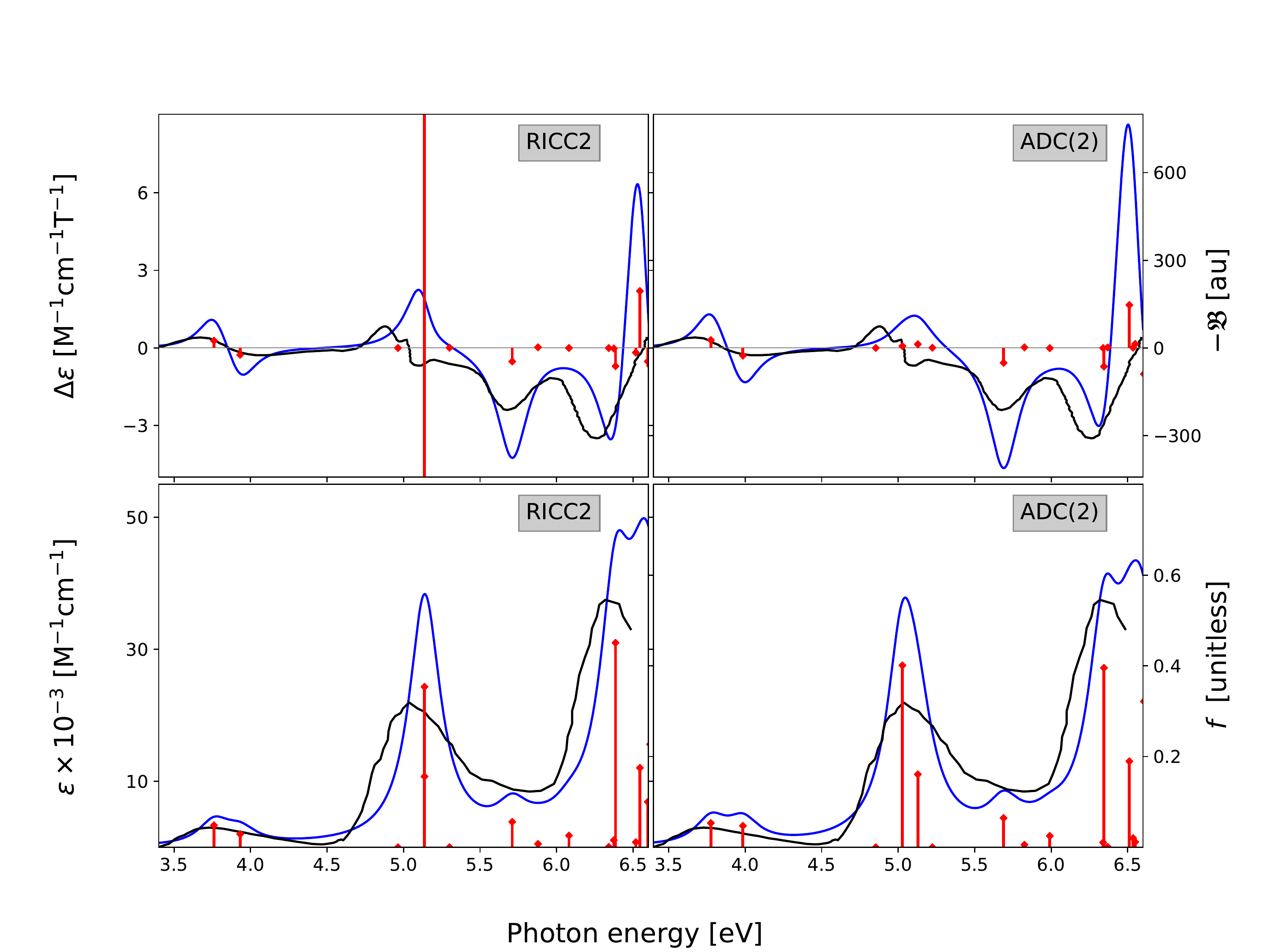}
\caption{1,4-naphthoquinone (NQ). 
MCD (upper panels) and OPA spectra (lower panels) at RICC2 and ADC(2) level. 
Experimental spectra recorded in cyclohexane (black lines) were digitized from Ref.\citenum{exp:Anthra:Naphta} and scaled by 10$^{4}$. The simulated spectra were shifted  by $-$0.39 eV (RICC2) and $-$0.38 eV (ADC(2)) to align with experiment.
Only one of the two dark transitions [at $\sim$2.81 eV (440 nm) and 2.72 eV (455 nm)], predicted in 
Ref.~\citenum{exp:Anthra:Naphta}, {was found at $\sim$2.5 eV (ADC(2)) and $\sim$2.6 eV (RICC2) after shifting.} 
This dark transition is not shown.
\label{fig:1-4-Naph:ADC2:CC2:shift}}
\end{figure*}

%AQ
{The experimental spectrum of AQ presents two positive features in the low-energy region, located 
%at $\sim$320 nm and at $\sim$270 nm},
at $\sim$3.85 eV and at $\sim$4.6 eV},
followed by a relatively intense pseudo-{$\mathcal{A}$} band in between 4.6 and 5.6 eV.
%260 and 220 nm. 
% \revS{QUESTION: are you enhancing certain regions in the spectrum when you plot? Those two features do not LOOK tiny in the plot! IF you enhanced them, then you should mention it here and clarify it in the figure too...}\revD{I do not.}
{Both SAOP/QZ3P1D and BP86/DZP TD-DFT calculations~\cite{Seth2008:b} reproduce {the first} feature in the low-energy part of the spectrum. 
However, they both predict an opposite sign pattern with respect to the experiment
for the pseudo-{$\mathcal{A}$} band. }
In contrast, RICC2 and ADC(2) yield the correct sign pattern of the intense pseudo-{$\mathcal{A}$} band, with positive and negative lobes due to the 2B$_{3u}$ and 2B$_{2u}$ transitions, respectively. They also reproduce the first weak feature, though shifted (it is located at $\sim$3.9 eV
%$\sim$300 nm
after the overall realignment with experiment). 
% %and the peak at $\sim$300 nm. 
% \revD{In both spectra \revS{WHICH SPECTRA ARE WE NOW TALKING ABOUT?}
% the position of the tiny band in the middle predicted to be in agreement with experiment. 
%WE ALREADY WROTE THAT TDFT  YEALD }
Both second-order methods fail to predict sign and intensity of the feature around 4.6 eV: 
%270 nm: 
for RICC2 this transition is basically dark to MCD, while for ADC(2) it is negative (it has a small positive $\mathcal{B}$).
{From the results reported in Ref.~\citenum{Seth2008:b}, we could not
discern whether this feature was reproduced or not by the SAOP/QZ3P1D and BP86/DZ methods.}
%Memorizing, that the experimental spectrum was recorded in cyclohexane
Thus, none of the simulated MCD spectra is fully consistent with the experimental MCD spectrum of AQ. Nonetheless, both RICC2 and ADC(2) reproduce the most important spectral features of AQ. 

\begin{figure*}[hbt!]
\centering
\includegraphics[width=\linewidth]{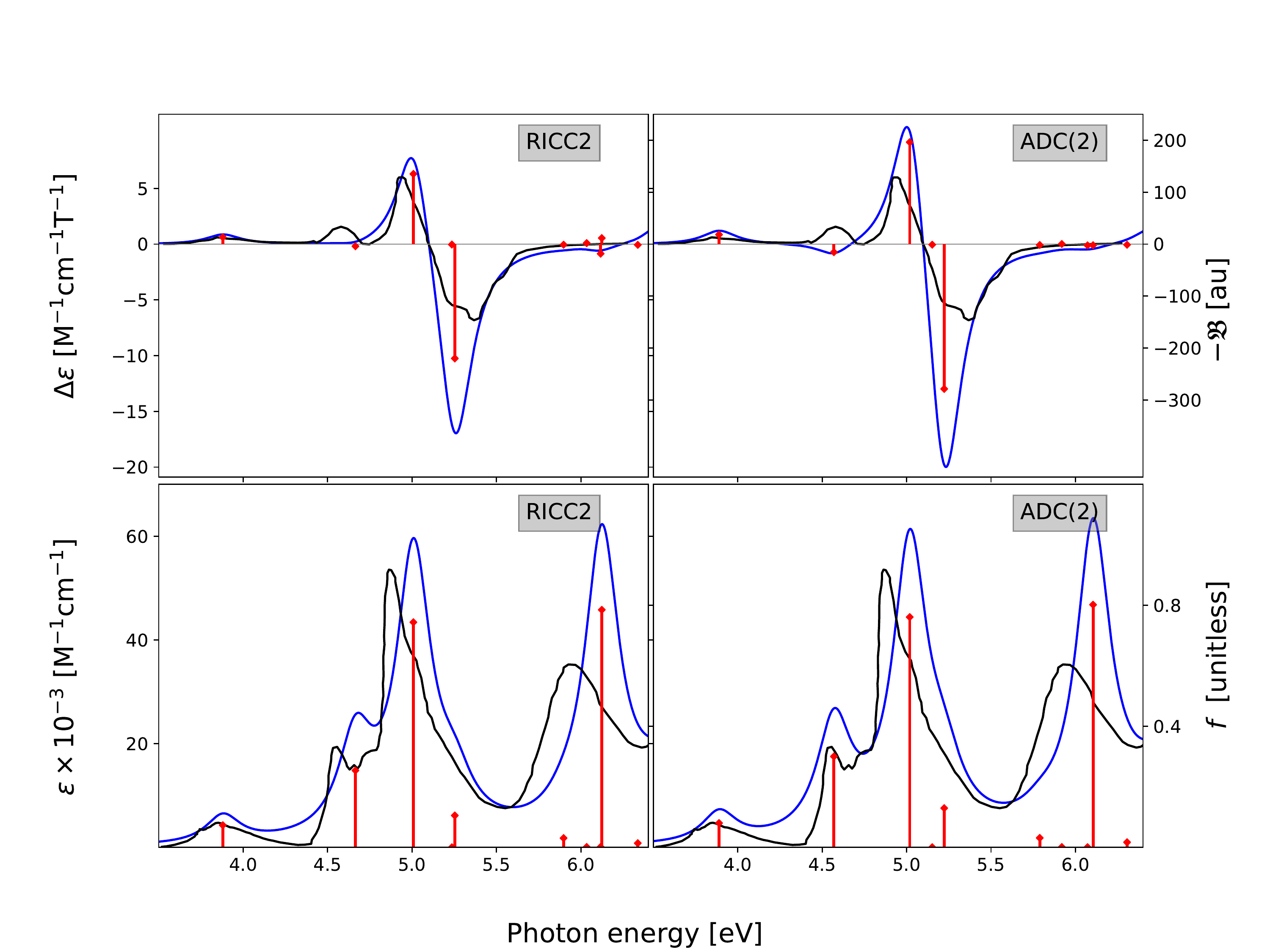}
\caption{
% \revS{QUESTION:are you enhancing certain regions in the spectrum when you plot? IF YES, then you should mention it and show it in the figure too...}
9,10-anthraquinone (AQ). MCD (upper panels) and OPA spectra (lower panels) at the RICC2 and ADC(2) level. Experimental spectra recorded in cyclohexane are presented as black lines, digitized from Ref.\citenum{exp:Anthra:Naphta} and scaled by 10$^{4}$. The simulated spectra were shifted 
by $-$0.46 eV to align 
{with the inflection point of the pseudo-{$\mathcal{A}$} band in the experiment}. 
None of the two dark transitions (at around 3.31 eV (375 nm) and 2.95 eV (420 nm)) otherwise predicted in Ref.~\citenum{exp:Anthra:Naphta}, was found. 
\label{fig:9-10Anthra:ADC2:CC2:shift}}
\end{figure*}

The experimental MCD spectrum of 1NA was recorded in cyclohexane by~\citeauthor{exp:1NA}.~\cite{exp:1NA} It includes a bisignate feature in the low-energy region of the spectrum and two negative peaks in the region $\sim$3.4-6 eV. 
%$\sim$260-210 mn. 
A computational study at the RICC2,
TD-B3LYP and TD-CAM-B3LYP levels of theory was recently reported by~\citeauthor{ghidinelli_1-Naph}~\cite{ghidinelli_1-Naph}
% All simulated spectra are consistent with the experimental one but slightly overestimate its intensity. 
\citeauthor{ghidinelli_1-Naph} noted that RICC2 and TD-CAM-B3LYP methods well reproduced the profile of the experimental spectrum, while TD-B3LYP yielded 
a small positive peak in-between the two negative features in the high-energy region. 
Close inspection of our results reveals very similar results for RICC2 and ADC(2), though one of the negative  $\mathcal{B}$ terms in the region around %220 nm 
$\sim$6.1 eV
is slightly larger for ADC(2) than 
RICC2. If a $\gamma$ value of 1000 cm$^{-1}$ is used for the broadening, this results in the emergence of a small  positive band for ADC(2), like it was observed for TD-B3LYP.\cite{ghidinelli_1-Naph} One should not forget, however, that the appearance of the spectrum depends on the chosen value of the broadening.
{Adjusting $\gamma$ to 5.7$\times$10$^{-3}$ a.u., the broadened ADC(2) spectrum becomes fully consistent with the RICC2 one.}
For both second-order methods, the bisignate feature is due to the $L_a$ and $L_b$ transitions: 1A' (negative band) and 2A' (positive band). Several transitions, most of them of relatively low-intensity
contribute to the first negative band, while the second one is, in both methods, due to the 7A' transition. 

\begin{figure*}[hbt!]
\centering
\includegraphics[width=\linewidth]{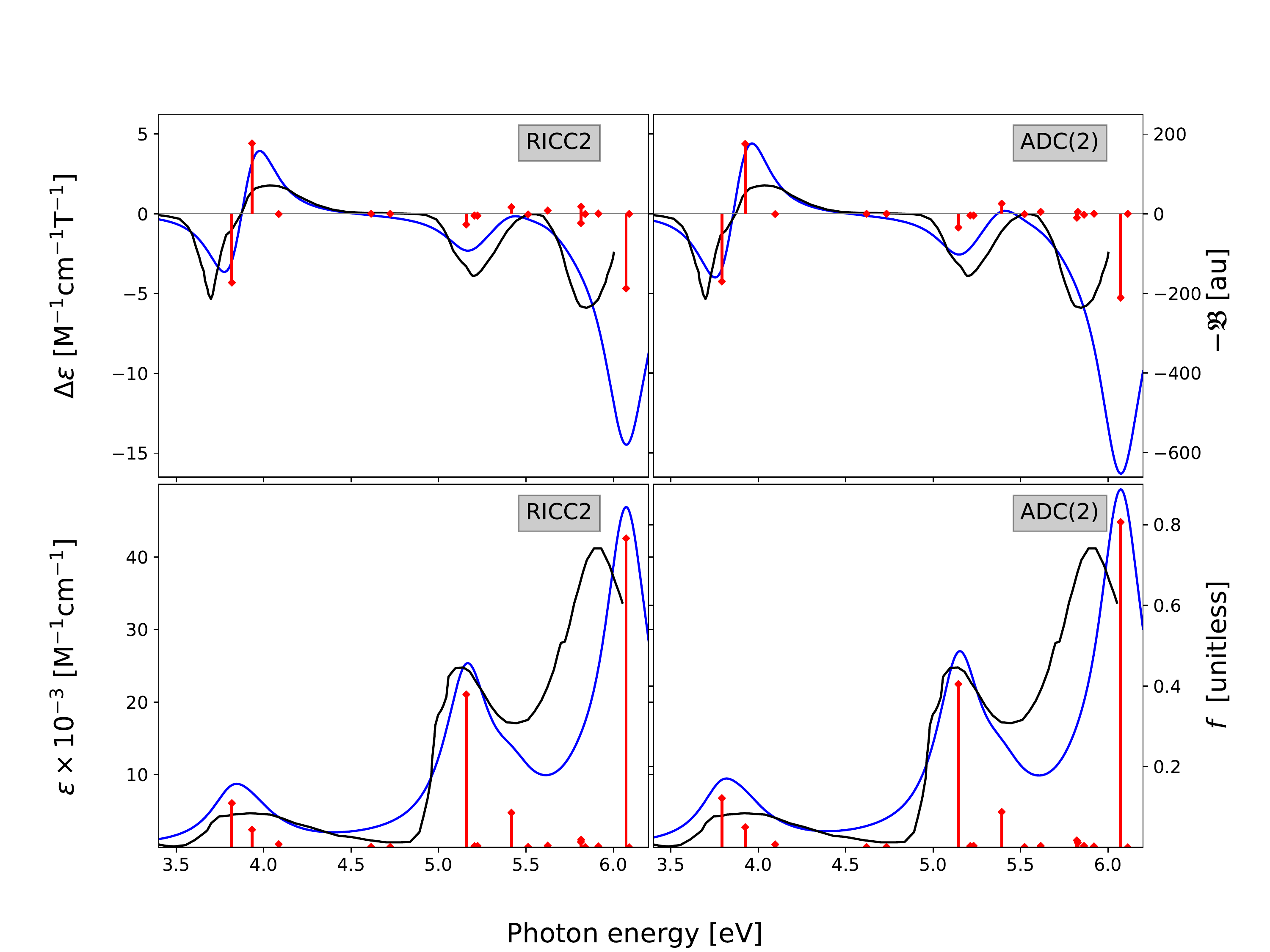}
\caption{1-Naphthylamine (1N). MCD (upper panels) and OPA spectra (lower panels) at RICC2 and ADC(2) level. Experimental spectrum, recorded in cyclohexane, is presented as black lines, digitized from Ref.\citenum{ghidinelli_1-Naph}. Both simulated spectra were shifted by $-$0.17 eV to align with the lower-energy experimental inflection point. 
{Here, $\gamma$ = 5.7$\times$10$^{-3}$ a.u. was used.}
% \revD{The embedded plot represents the broaden ADC(2) spectrum with $\gamma$ value equal to 6.556$\times$10$^{-3}$.}
\label{fig:1-Naph:ADC2:CC2:shift}}
\end{figure*}

\clearpage
\section{Conclusions}
\label{Conclusions}
We have presented the derivation and implementation of a computational scheme for the $\mathcal{B}$ term of Magnetic Circular Dichroism within the ADC/ISR framework, thus extending the pool of spectroscopic effects that can be simulated using the ADC($n$) methods.  
In the lowest-energy region of the spectra, an overall qualitative agreement with experiment is observed for all ADC methods. 
%Though, we discourage the use of the ADC(2)-x, and it always requires relatively large realignment shifts.
{Modelling the higher energy region with the presented approach is more demanding as the density of electronic states increases and a large number of converged states is required. An alternative \textit{ansatz} is offered by the complex-polarization propagator approach which is more suited for that situation.}
% At higher energies, the situation becomes less clear, as the density of electronic states increases, and we cannot be sure to have converged a sufficient number of states; the complex-polarization propagator approach would probably be more convenient to use.
Work in this direction is currently ongoing.
%\revD{Our results show an agreement between ADC(2) and RICC2 methods in the transitions, responsible for the formation of the main spectral features.}

\section*{Supplementary Material}
% \revS{Must add something on what is in the SI}
Tables of raw spectral data (energies, oscillator strengths and (negative) ${\mathcal B}$ terms; ADC(2)-x spectra; ADC(2) natural transition orbitals of the first two MCD bands in  2-thiouracil; Cartesian coordinates of all molecules.
%\revD{
%Optimized structures and negative MCD $\mathcal{B}$ term computed at ADC(2), RICC2, ADC(3), as well as ADC(2)-x levels of theory are presented in the supplementary file. 
%We also present NTOs for the two first the most intense transitions of 2-thiouracil.
%} 

\section*{Acknowledgements}
This work was carried out with support from the European Union's Horizon 2020 Research and Innovation Programme under the Marie Sk{\l}odowska-Curie European Training Network, grant no. 765739 ({COSINE}).
S. C. acknowledges the Independent Research Fund Denmark--Natural Sciences, DFF-RP2 Grant No. 7014-00258B.
{We thank Dr. Xiaosong Li (University of Washington) for sending us the Cartesian coordinates of 1,4-naphthoquinone. S. C. acknowledges discussion with Dr. S. Ghidinelli (University of Brescia) and with Dr. Christof H{\"a}ttig (Ruhr Universit{\"a}t Bochum).}

\section*{Author declarations}
\subsection*{Conflict of Interest}
The authors have no conflicts to disclose.
\subsection*{Author contributions}
S. C. and A. D. conceptualized and supervised the project.
D. A. F., M. S. and D. R. R. derived the working equations and implemented the computational methodology. D. A. F. performed the calculations.
D. A. F. and S. C. drafted the manuscript. All authors revised the draft and discussed its scientific content.

\section*{Data Availability}
The data that supports the findings of this study are available within the article and its supplementary material. 

\clearpage
%\bibliographystyle{jcp}
%\bibliography{mcdadc.bib}
\bibliography{feniil_lib4tex}
\end{document}